\documentclass[10pt]{iopart}
\usepackage{graphicx,rotating,subfigure} 
\usepackage{iopams}  
\usepackage{dsfont}

\newcommand{\im}{\mbox{i}}
\newcommand{\be}{\begin{eqnarray}}
\newcommand{\ee}{\end{eqnarray}}
\def\refeq#1{(\ref{#1})}
\def\d{d}
\def\wt{\widetilde}
\def\nn{\nonumber}
\def\i{\int_{-\infty}^{\infty}}
\def\ip{\int_{0}^{\infty}}

\def\G{\Gamma}

\def\la{\lambda}
\def\g{\gamma}
\def\al{\alpha}

\def\eps{\epsilon}

\def\l{\left}
\def\r{\right}
\def\e{\mbox{e}}

\def\PhR{\phi_R}
\def\PhL{\phi_L}
\def\te{\mbox{e}}

\def\down{\downarrow}

\begin{document}
\title[The open $XXZ$-chain]{The open $XXZ$-chain: Bosonisation, Bethe ansatz and logarithmic corrections}
\author{Jesko Sirker}
\address{Department of Physics and Astronomy, University of British Columbia,
Vancouver, B.C., Canada V6T 1Z1}
\author{Michael Bortz}
\address{Department of Theoretical Physics, Research School of Physics and Engineering, Australian National University, Canberra ACT 0200, Australia}
\begin{abstract}
  We calculate the bulk and boundary parts of the free energy for an open
  spin-$1/2$ $XXZ$-chain in the critical regime by bosonisation. We identify
  the cutoff independent contributions and determine their amplitudes by
  comparing with Bethe ansatz calculations at zero temperature $T$. For the
  bulk part of the free energy we find agreement with Lukyanov's result
  [Nucl.~Phys.~B {\bf 522}, 533 (1998)]. In the boundary part we obtain a
  cutoff independent term which is linear in $T$ and determines the
  temperature dependence of the boundary susceptibility in the attractive
  regime for $T\ll 1$. We further show that at particular anisotropies where
  contributions from irrelevant operators with different scaling dimensions
  cross, logarithmic corrections appear. We give explicit formulas for these
  terms at those anisotropies where they are most important. We verify our
  results by comparing with extensive numerical calculations based on a
  numerical solution of the $T=0$ Bethe ansatz equations, the finite
  temperature Bethe ansatz equations in the quantum-transfer matrix formalism,
  and the density-matrix renormalisation group applied to transfer matrices.
\end{abstract}
\section{Introduction}
\label{intro}
The spin-$1/2$ $XXZ$-chain is described by the Hamiltonian  
\begin{equation}
\label{B1}
H = J\sum_{j=1}^{N,N-1}\left[\frac{1}{2}\left(S^+_jS^-_{j+1}+S^-_jS^+_{j+1}\right)+\Delta
  S^z_jS^z_{j+1}\right] \; .
\end{equation}
Here $J>0$ is the coupling constant and $N$ is the number of sites. The model
is critical for anisotropy $-1\leq\Delta\leq 1$. If we take $N$ as upper
boundary in the sum in Eq.~(\ref{B1}) we have periodic boundary conditions
(PBC) whereas taking the sum only up to $N-1$ corresponds to open boundary
conditions (OBC).

The reasons for the popularity of this model are twofold: On the one hand,
this model does indeed capture the basic physics in some real physical systems
as for example SrCuO$_3$ \cite{MotoyamaEisaki}. On the other hand, the model
is exactly solvable by Bethe ansatz. Furthermore, in the continuum limit at
low energies it is equivalent to a free boson model up to irrelevant
operators. The exact solution at zero temperature $T$ for the isotropic
antiferromagnetic case $\Delta=1$ and PBC has been first constructed by Bethe
\cite{Bethe}. This so called Bethe ansatz (BA) has later been generalised to
allow also for the calculation of ground-state properties in the anisotropic
case \cite{YangYang,YangYang2}.  Thermodynamic properties can also be studied
by using either the thermodynamic Bethe ansatz (TBA)
\cite{tak99} or the quantum transfer-matrix (QTM) approach
\cite{SuzukiInoue,Kluemper_HB}.  However, at finite temperature even the
equations obtained for simple thermodynamic quantities like free energy,
specific heat or susceptibility can often only be solved numerically.

The BA solution for the $XXZ$-chain with OBC at $T=0$ has been constructed by
Gaudin \cite{gau83}, Alcaraz {\it et al.} \cite{AlcarazBarber}, and Sklyanin
\cite{Sklyanin}. The application of the TBA for this case seems to be
difficult or even impossible as is discussed in more detail in
\cite{BortzSirker,GoehmannBortz}. A modification of the QTM approach is
possible as demonstrated in \cite{GoehmannBortz} but an evaluation of the
obtained formulas still seems to be a formidable task.

An entirely different approach is based on bosonisation (see
\cite{Affleck_lesHouches} and references therein). Here a low-energy effective
theory is derived which is just a free boson model up to irrelevant operators.
The main advantage of this approach compared to the BA is that any
thermodynamic quantity and also any kind of correlation function can be
calculated very easily for the free boson model. Corrections to this free
boson approximation are obtained by doing perturbation theory in the
irrelevant operators. A disadvantage is that the spin-wave velocity $v$, the
Luttinger parameter $K$ as well as the amplitudes of the irrelevant terms in
the Hamiltonian can only be obtained perturbatively for $|\Delta|\ll 1$.
However, one can use the BA to fix $v$ and $K$ which in the language of the
bosonic model is equivalent to summing exactly an infinite series of terms
which would renormalise the bare values of $v$ and $K$. Lukyanov has
demonstrated that the BA can also be used to fix the amplitudes of the leading
irrelevant operators in the bosonic model \cite{Lukyanov}. This allows to give
analytic formulas at low temperatures for the free energy and derived
quantities. It also allows to calculate the correlation amplitudes for the
leading and some sub-leading terms in an asymptotic expansion of spin-spin
correlation functions \cite{LukyanovTerras}.

The $XXZ$-chain with open boundaries has attracted considerable attention
recently
\cite{FujimotoEggert,FurusakiHikihara,BortzSirker,SirkerBortz,GoehmannBortz}
because it is the simplest model for a spin-chain containing a small number of
non-magnetic impurities which cut the chain into parts with essentially free
boundaries. It has been shown that the leading irrelevant term in the bosonic
model for anisotropy $1/2<\Delta\leq 1$ leads to a contribution in the
boundary susceptibility $\chi_B$ ($\mathcal{O}(1)$ part of the susceptibility
if the bulk part is $\mathcal{O}(N)$) which diverges as $T^{2K-3}$ where the
Luttinger parameter $K$ varies between $K=3/2$ at $\Delta=1/2$ and $K=1$ at
$\Delta=1$. The $1/T$ behaviour (up to a logarithmic correction) at the
isotropic point is of special interest because it is a Curie-like contribution
without any free spins being present. The isotropic ferromagnetic point has
also been studied recently \cite{SirkerBortz} and it has been found that the
boundary susceptibility $\chi_B\sim -1/T^3$, i.e., the boundary susceptibility
diverges more rapidly and with opposite sign than the bulk susceptibility
$\chi\sim 1/T^2$.

In this article we want to use Lukyanov's method \cite{Lukyanov} to obtain
explicit expressions for the free energy of the open $XXZ$-chain in the entire
critical regime $-1<\Delta <1$. To do so we will apply perturbation theory in
the leading irrelevant operators of the bosonic theory and compare with
results obtained by Bethe ansatz. This will allow us to fix the amplitudes of
the irrelevant terms. Our result for the bulk part of the free energy will
confirm Lukyanov's result, however, we will obtain in addition the boundary
part.  In particular, we find that the boundary susceptibility in the
attractive regime behaves as $\chi_B\sim\mbox{const}+T$ compared to the
$\chi_B\sim \mbox{const}+T^{2K-3}$ behaviour found before for $0 <\Delta <1$
\cite{FujimotoEggert,FurusakiHikihara,BortzSirker}. Another focus of our work
are logarithmic corrections. We will find that the amplitudes in the
low-temperature expansion of the free energy as well as the amplitudes in the
expansion of the ground-state energy in terms of the magnetic field $h$ show
divergences at infinitely many anisotropies. We will show that these
divergences occur because contributions from terms with different scaling
dimensions cross at these points. The divergences in the two terms which cross
cancel each other leading to a logarithmic correction. We will give explicit
formulas for these corrections at those anisotropies where they are most
important.

Our paper is organised as follows: In section \ref{Bos} we bosonise the open
$XXZ$-chain and expand the free energy in powers of $h,T$ by using
perturbation theory in the leading irrelevant operators. In section \ref{BA}
we present the zero temperature Bethe ansatz solution for this model. In
particular, we show how the Wiener-Hopf method can be used to obtain leading
and next-leading terms in an expansion of the ground-state energy in powers of
$h$. 
We then determine the amplitudes of the irrelevant operators in the field
theory approach in section \ref{Ampl} by comparing with BA. We confirm
Lukyanov's result obtained by the same approach for the periodic case. In
section \ref{main} we summarise our result for the bulk and boundary part of
the susceptibility. We analyse where the amplitudes of the different terms
diverge and give explicit formulas for the logarithmic corrections caused, at
points where they appear in the leading or sub-leading term. We check all our
analytical results by comparing with a number of numerical data obtained by a
numerical solution of the $T=0$ BA equations, the BA equations in the QTM
approach, and the density-matrix renormalisation group (DMRG) applied to
transfer matrices in section \ref{numerics}. In the final section we give a
short summary and conclusions.
\section{Bosonisation for the open chain}
\label{Bos}
First, we want to briefly review the bosonisation procedure for the
$XXZ$-chain (see for example \cite{Affleck_lesHouches}) to clarify our
notation and to remind the reader how a priori unknown parameters in the
bosonic model can be determined by Bethe ansatz.
 
Using the Jordan-Wigner transformation we can represent the $XXZ$-chain with
OBC as given in Eq.~(\ref{B1}) in terms of fermionic operators $\psi_j$:
\begin{eqnarray}
\label{B2}
&& H = H_0 + H_{\mbox{\small int}} \\
 &&=J\sum_{j=0}^{N}\left[\frac{1}{2}\left(\psi^\dagger_j \psi_{j+1}+\psi^\dagger_{j+1}\psi_j
    \right)+\Delta \left(\psi^\dagger_j \psi_j
  -\frac{1}{2}\right)\left(\psi^\dagger_{j+1} \psi_{j+1} -\frac{1}{2}\right)\right] \nn
\end{eqnarray}
where we have added two sites and impose the boundary conditions
$\psi_0=\psi_{N+1}=0$. For $\Delta=0$ we are left with a free-fermion
Hamiltonian $H_0$. The low-temperature properties of this model can be studied
by linearising the dispersion around the Fermi points $\pm k_F$ and going to
the continuum limit. To do so, we introduce right and left moving fermions by
\begin{equation}
\label{B3}
\frac{\psi_j}{\sqrt{a}} \approx \e^{\rmi k_F x} \psi_R(x) + \e^{-\rmi k_F x} \psi_L(x) 
\end{equation}
with $x=ja$ and $a$ being the lattice constant. In the half-filled case we
have $k_F = \pi/2a$ (no magnetic field applied in (\ref{B1})). The continuum
limit is achieved by approximating $\psi_{j+1}=\psi(x)+a\partial_x\psi(x)$. This
leads in lowest order in $a$ to the continuum Hamiltonian
\begin{equation}
\label{B4}
H_0 = \im v_0 \int_0^L dx \left[ \psi_R^\dagger \partial_x\psi_R
  -\psi_L^\dagger \partial_x\psi_L \right] 
\end{equation}
where $L=Na$ and the spin-wave velocity is given by $v_0=Ja$. 

This Hamiltonian can be bosonised by 
\begin{equation}
\label{BosFormula}
\psi_R=\frac{1}{\sqrt{2\pi}}\exp(\rmi\sqrt{4\pi}\phi_R) \quad ; \quad \psi_L=\frac{1}{\sqrt{2\pi}}\exp(-\rmi\sqrt{4\pi}\phi_L) 
\end{equation}
where $\phi = \phi_R + \phi_L$ is a bosonic field obeying the standard
commutation rule $[\phi(x),\Pi(x')]=\rmi\delta(x-x')$ with
$\Pi=v_0^{-1}\partial_t\phi$. Up to total derivatives the Hamiltonian (\ref{B4})
takes the following form in terms of the bosonic fields
\begin{equation}
\label{B7}
H_0 = v_0 \int_0^L dx \left[ (\partial_x\phi_R)^2+(\partial_x\phi_L)^2\right]=\frac{v_0}{2} \int_0^L dx \left[\Pi^2+(\partial_x\phi)^2\r] \; . 
\end{equation}
The interaction part of the Hamiltonian (\ref{B3}) can also be expressed in
terms of the bosonic fields
\begin{eqnarray}
\label{B14}
H_{\mbox{\footnotesize int}} &=& J\Delta\sum_j \l(:\Psi^\dagger_j\Psi_j:\; :\Psi^\dagger_{j+1}\Psi_{j+1}:\r)
\nn \\
&\approx& \frac{2\Delta v_0}{\pi} \int dx \l[(\partial_x\phi_R)^2+(\partial_x\phi_L)^2+2(\partial_x\phi_R)(\partial_x\phi_L)\r] \; .
\end{eqnarray}
The first two terms are identical to the free part (\ref{B7}) and yields only a
renormalisation of the velocity
\begin{equation}
\label{B20}
 v_0\rightarrow v=v_0(1+2\Delta/\pi) \; . 
\end{equation}
The third term is an interaction between right and left fields. Crucially,
even with this term included we still obtain a free boson model in terms of
the fields $\phi,\Pi$
\begin{equation}
\label{B20.1}
H=H_0+ H_{\mbox{\footnotesize int}} =\frac{v}{2}\int dx \l[(1-c)\Pi^2+(1+c)(\partial_x\phi)^2\r] 
\end{equation}
with $c=2\Delta/(\pi+2\Delta)$. We now rescale the fields by
\begin{equation}
\label{B18}
\phi \rightarrow \frac{\phi}{\sqrt{4\pi}R} \; ; \; \Pi \rightarrow \sqrt{4\pi}R
\,\Pi \; ; \; R^2 = \frac{1}{4\pi}\sqrt{\frac{1+c}{1-c}}=\frac{1}{4\pi}\sqrt{1+\frac{4\Delta}{\pi}} 
\end{equation}
and define the Luttinger parameter
\begin{equation}
\label{B21}
K = \frac{1}{2\pi R^2} = 2\l(1+\frac{4\Delta}{\pi}\r)^{-1/2}\approx 2-\frac{4\Delta}{\pi} \; . 
\end{equation}
The bosonised version of the Hamiltonian (\ref{B1}) is then given by
\begin{equation}
\label{BosHam}
H =\frac{v}{2} \int_0^L dx \left[\Pi^2+(\partial_x\phi)^2\r] \; . 
\end{equation}

How serious should one take the values for $v,K$ given in (\ref{B20}) and
(\ref{B21}), respectively? We have derived an effective bosonic Hamiltonian
for the spin chain by expanding all terms up to linear order in the lattice
constant $a$. So clearly, there are higher order corrections to the
Hamiltonian (\ref{BosHam}). In fact, there are infinitely many irrelevant
terms which have to be added to this bosonic Hamiltonian to obtain an accurate
description of the spin chain. Some of these terms (in fact, also infinitely
many) will further renormalise the values of $v,K$. So
Eqs.~(\ref{B20},\ref{B21}) are {\it perturbative} results for $v,K$ expected
to be valid only if $\Delta\ll 1$. On the other hand, the model is exactly
solvable by BA and we will discuss this solution in section \ref{BA}. Among
other things, BA allows to calculate the spin-wave velocity exactly. In other
words, the BA allows to sum up exactly the infinite perturbative series for
$v$ obtained by bosonisation. In section \ref{BA} we will also see that the
leading term in the bulk susceptibility at zero temperature is a function of
$v$ and $K$ only. This allows it to determine also $K$ exactly and leads to
the well known results
\begin{eqnarray}
\label{B22}
v &=& \frac{\pi\sqrt{1-\Delta^2}}{2\arccos\Delta} \\
K &=& \frac{1}{2\pi R^2} =\frac{\pi}{\pi-\arccos\Delta} \; .
\end{eqnarray}
Note that these exact formulas agree with (\ref{B20},\ref{B21}) in lowest
order in $\Delta$ as expected. When we use the bosonised Hamiltonian with
these values for $v,K$ we should keep in mind that all contributions from
perturbation theory in the irrelevant operators leading to a renormalisation
of these parameters are already accounted for.

Finally, we want to give the direct relations between the bosonic and the spin
operators \cite{Affleck_lesHouches} for later use 
\begin{eqnarray}
\label{B23}
S^z_j&\approx& \frac{1}{2\pi R}\partial_x\phi+(-1)^j \mbox{const}
\,\cos\frac{\phi}{R} \nn \\
S^{-}_j&\approx&\exp(2\pi\rmi R\tilde{\phi})\l(\mbox{const}\, \cos\frac{\phi}{R}
+(-1)^j \mbox{const}\r) \; .
\end{eqnarray}
Here $\tilde{\phi} = \phi_L-\phi_R$ is the dual field and we clearly have to
identify $\phi = \phi +2\pi R$ and $\tilde{\phi}=\tilde{\phi}+1/R$, i.e.,
$\phi,\tilde{\phi}$ are compact fields.
\subsection{The spin-chain in a magnetic field and the mode expansion for OBC}
\label{Mag}
When we add a magnetic field term $H_M = h\sum_{j=1}^N S^z_j$ to the
Hamiltonian (\ref{B1}), the bosonic Hamiltonian (\ref{BosHam}) has to be replaced
by
\begin{equation}
\label{B28}
H = \frac{v}{2} \int_0^L dx \left[\Pi^2+(\partial_x\phi)^2 -\frac{h}{\pi Rv}\partial_x\phi\r]  
\end{equation}
according to Eq.~(\ref{B23}). By performing a shift in the boson field 
\begin{equation}
\label{B28b}
\phi\rightarrow \phi+hx/(2\pi R v)
\end{equation}
we can rewrite this as
\begin{equation}
\label{B29}
H = \frac{v}{2} \int_0^L dx \left[\Pi^2+(\partial_x\phi)^2\r]
  -L\frac{K h^2}{4\pi v}  \; . 
\end{equation}
The bulk susceptibility in lowest order is therefore given by
$\chi_{\mbox{\tiny bulk}} = K/2\pi v$. The magnetic field shifts the Fermi
points away from $k_F=\pi/2$ so that according to Eq.~(\ref{B3}) the right-
and left-moving fermions are no longer situated near these points. It is,
however, easy to see that the shift in the Bose field (\ref{B28b}) exactly
compensates for this shift so that the fermion fields $\Psi_R,\Psi_L$
belonging to the shifted Bose field have again $k_F=\pi/2$. The boundary
conditions for the field $\phi$ in (\ref{B29}) are therefore exactly the same
as in a system without magnetic field.

We first consider the boundary conditions for the $\psi$-field:
\begin{eqnarray}
\label{B31}
\psi(0)&=&\psi(N+1)= 0 \nn \\
&\Rightarrow & \psi_L(0) + \psi_R(0) = 0 \nn \\
&\Rightarrow & \psi_L(L) +(-1)^{N} \psi_R(L) = 0
\end{eqnarray}
where $L=Na$ and the last two relations have been derived from (\ref{B3}).
{}From the bosonisation formula (\ref{BosFormula}) and the rescaling relation
(\ref{B18}) it follows that $[\PhR(x),\PhL(x)] = \im\pi R^2$ at any site
$x\neq 0,L$, i.e., the left and right fields do not commute in general. This
is, however, different at the boundaries where $\PhR,\PhL$ are related
according to (\ref{B31}) leading to $[\PhR(0),\PhL(0)] = [\PhR(L),\PhL(L)]=0$
and
\begin{equation}
\label{B33}
 \PhR(0) + \PhL(0) = \pi R+2\pi R n \; , \; \PhR(L) + \PhL(L) = 2\pi R n' 
\end{equation}
where $n$ is an integer. $n'$ is a half-integer if $N$ is even and an integer
if $N$ is odd. Eq.~(\ref{B23}) allows us to fix the numbers $n,n'$ because
$S^z_{\mbox{\tiny tot}}= \sum_j S^z_j \approx [\phi(L)-\phi(0)]/2\pi R$. The total spin $S^z_{\mbox{\tiny tot}}$ should be an integer for $L$ even and a
half-integer for $L$ odd. Fixing the boundary condition for $\phi(0)$
we therefore obtain
\begin{equation} 
\label{B45}
\phi(0) = \pi R \quad ; \quad  \phi(L) = \pi R +2\pi R
S^z_{\mbox{\tiny tot}} \; .
\end{equation}
This leads to the following mode expansion for OBC \cite{EggertAffleck92}  
\begin{eqnarray}
\label{B50}
\fl \qquad\phi(x,t) = \pi R +2\pi R S^z_{\mbox{\tiny tot}}\frac{x}{L} + \sum_{n=1}^N
\frac{\sin\l(\pi nx/L\r)}{\sqrt{\pi n}}\l(\e^{-\rmi\pi n\frac{vt}{L}}a_n + \e^{\rmi\pi n\frac{vt}{L}}a_n^\dagger\r),     
\end{eqnarray}
where $a_n$ is a bosonic annihilation operator. 
\subsection{Correlation functions for the free boson model}
\label{corr}
{}From the mode expansion (\ref{B50}) we can directly calculate the basic
correlation functions. As we are not interested in finite size effects but
rather in the bulk and boundary properties of the chain in the thermodynamic
limit we will consider $L\rightarrow\infty$. The results are then valid for
the semi-infinite line, i.e.~, a system with only one boundary.

We start with the basic correlation function
\begin{eqnarray}
\label{B52}
\fl \langle\phi(x_1,\tau_1)\phi(x_2,\tau_2)\rangle_{\mbox{\tiny OBC}} \nn \\
\fl \qquad\stackrel{L\rightarrow\infty}{\longrightarrow}
-\frac{1}{4\pi}\l(\ln[(x_1-x_2)^2+v^2(\tau_1-\tau_2)^2]-\ln[(x_1+x_2)^2+v^2(\tau_1-\tau_2)^2]\r)
\nn \\
\fl\qquad= \frac{1}{4\pi}\l(\ln(\bar{z}_1-z_2)+\ln(z_1-\bar{z}_2)-\ln(z_1-z_2)-\ln(\bar{z}_1-\bar{z}_2)\r)
\end{eqnarray}
where $\tau$ represents the imaginary time. In the last line we have
introduced new coordinates 
\begin{eqnarray}
\label{B53}
z &=& -\im(x-vt) = v\tau -\im x \quad , \quad
\bar{z} = \im(x+vt) = v\tau +\im x \; .
\end{eqnarray}
For PBC we would find instead
\begin{eqnarray}
\label{B55}
 &&\langle\phi(x_1,t_1)\phi(x_2,t_2)\rangle_{\mbox{\tiny PBC}} \nn \\
&\stackrel{L\rightarrow\infty}{\longrightarrow}& -\frac{1}{4\pi}\ln[(x_1-x_2)^2+v^2(\tau_1-\tau_2)^2]+\mbox{const} \\
&=& -\frac{1}{4\pi}\l(\ln(z_1-z_1) + \ln(\bar{z}_1-\bar{z}_2)\r)
+\mbox{const} \; . \nn 
\end{eqnarray}
{}From these basic correlation functions one can easily obtain correlation
functions as for example $\langle\partial_x\phi\partial_x\phi\rangle$ by
taking partial derivatives.  The correlation function for the exponential
fields can be obtained by
\begin{eqnarray}
\label{B59}
\fl\l\langle\exp\l(\pm\im\frac{2\phi(x,t)}{R}\r)\r\rangle =\sum_{n=0}^\infty
\frac{\l\langle\l(\pm\im\frac{2\phi}{R}\r)^n\r\rangle}{n!} = \sum_{n=0}^\infty
\frac{(-4/R^2)^n}{(2n)!}\langle\phi^{2n}\rangle   \nn \\
\fl \qquad = \sum_{n=0}^\infty \frac{(-4/R^2)^n}{(2n)!} \frac{(2n)!}{2^nn!}
\l(\langle\phi^{2}\rangle\r)^n = \exp\l(-\frac{2}{R^2}\langle\phi(x,t)\phi(x,t)\rangle\r) 
\end{eqnarray} 
and
\begin{eqnarray}
\label{B61}
\fl \l\langle\exp\l(\im\frac{2\phi(x_1,\tau_1)}{R}\r)\exp\l(\pm\im\frac{2\phi(x_2,\tau_2)}{R}\r)
\r\rangle =\l\langle\exp\l(\im\frac{2}{R}(\phi(x_1,\tau_1)\pm\phi(x_2,\tau_2))\r)\r\rangle
\nn\\
\fl \qquad =\exp\l(-\frac{2}{R^2}\langle (\phi(x_1,\tau_1)\pm\phi(x_2,\tau_2))^2\rangle\r)
 \\
\fl \qquad =\exp\l(-\frac{2}{R^2}[\langle\phi^2(x_1,\tau_1)\rangle +
\langle\phi^2(x_2,\tau_2)\rangle \pm 2
\langle\phi(x_1,\tau_1)\phi(x_2,\tau_2)\rangle ]\r) \; . \nn
\end{eqnarray}
These formulas for the correlation functions will allow us to calculate the
free energy of the model on a semi-infinite line perturbatively in subsection
\ref{pert}.
\subsection{The free fermion and the free boson model}
\label{freeE}
The free part $H_0$ of the Hamiltonian in Eq.~(\ref{B2}) can easily be solved
by Fourier transform. The free energy is given by
\begin{equation}
\label{f0}
 F=-T\sum_{n=1}^N\ln\l(1+\e^{-\beta\cos k_n}\r) \quad\mbox{with}\quad
 k_n=\frac{\pi}{(N+1)}n 
\end{equation}
where we set the lattice constant $a$ and the coupling $J$ equal to one. The
bulk and boundary part can be obtained by using the Euler-Maclaurin summation
formula
\begin{equation}
\label{f0.1}
\fl F=-\frac{T}{\pi}(N+1)\int_0^\pi dk\,\ln\l(1+\e^{-\beta\cos k}\r) +\frac{T}{2}\l(\ln(1+\e^\beta)+\ln(1+\e^{-\beta})\r)
\end{equation}
leading to the following low-temperature expansion 
\begin{eqnarray}
\label{f0.2}
F &=&
N\l(-\frac{1}{\pi}-\frac{\pi}{6}T^2-\frac{7\pi^3}{360}T^4+\mathcal{O}(T^6)\r) \\
&+&
\frac{1}{2}-\frac{1}{\pi}-\frac{\pi}{6}T^2-\frac{7\pi^3}{360}T^4+\mathcal{O}(T^6)+\mathcal{O}(T\e^{-1/T})
\; . \nn 
\end{eqnarray}
This means that bulk and boundary part have exactly the same temperature
dependence if $T\ll 1$. The same is also true for the bulk and boundary
susceptibility \cite{BortzSirker}
\begin{equation}
\label{f0.3}
 \chi_{\mbox{\tiny bulk}}(h=0) = \chi_B(h=0) =
 \frac{1}{\pi}+\frac{\pi}{6}T^2+\mathcal{O}(T^4) \; .
\end{equation}

Consider on the other hand the free boson model (\ref{BosHam}). Using the mode
expansion (\ref{B50}) the Hamiltonian takes the form
\begin{equation}
\label{f1}
 H=v\sum_{l=1}^N \frac{\pi l}{L}\l(a_l^\dagger a_l +\frac{1}{2}\r) \; .
\end{equation}
To identify the terms which are independent of the details of the dispersion
relation we introduce a momentum cutoff $\Lambda$. The free energy is then
given by
\begin{equation}
\label{f3}
F = \sum_{l=1}^{\Lambda N} \frac{v\pi l}{2L} +T \sum_{l=1}^{\Lambda
  N}\ln\l[1-\exp\l(-\frac{v\pi l}{TL}\r)\r] \; .
\end{equation}
The first part is the ground state energy $E_0$ which is clearly cutoff
dependent. We can rewrite the second part and obtain 
\begin{equation}
\label{f3.1}
F = E_0 -T \sum_{n=1}^{\infty}\frac{1}{n}\frac{1-\exp\l(-\frac{v\pi\Lambda}{aT} n\r)}{\exp\l(\frac{v\pi n}{TL}\r)-1} \; .
\end{equation} 
The exponential factor in the numerator is $\sim\mathcal{O}(\exp(-\Lambda/T))$
and can therefore be ignored if $T\ll 1$. If we now consider the limit
$N\rightarrow\infty$ with $T$ fixed we obtain
\begin{equation}
\label{f3.2}
F = E_0 -L\frac{T^2}{v\pi}\sum_{n=1}^\infty\frac{1}{n^2}=E_0-L\frac{\pi
  T^2}{6v} \; .
\end{equation} 
This means that the free energy of the continuum free boson model does not
contain any boundary terms. The reason for this is the continuum limit which
leads to $a(N+1)\rightarrow Na =L$. All boundary contributions in the field
theory therefore have to come from higher order corrections to the free boson
model.  We expect that the same powers of $T,h$ which are present in the
boundary free energy and the boundary susceptibility for the free fermion
model (which we will call regular terms afterwards) are also present for the
general case $-1<\Delta <1$ and only the coefficients of these terms will
change with anisotropy. By dimensional analysis it becomes clear that the
regular terms in boundary quantities obtained by perturbation theory in the
irrelevant operators have to involve the short distance lattice cutoff. It is
therefore impossible to determine the coefficients of these terms within the
field theory approach. We will come back to this point in the next section.
The only boundary terms which can be reliably obtained from the field theory
are therefore those which have no analogue in the bulk part. Such terms must
have an amplitude which vanishes at the free fermion point where bulk and
boundary parts show the same dependence on $T,h$ at low energies.
\subsection{Perturbation theory in the irrelevant operators}
\label{pert}
As we have already mentioned when deriving the bosonic Hamiltonian
(\ref{BosHam}) from the original spin-model (\ref{B1}) there are higher order
corrections.  These corrections are all irrelevant in the whole critical
regime except of the umklapp term which becomes marginal at $\Delta=1$. This
allows it to calculate their effects systematically by simple perturbation
theory.

The bosonised Hamiltonian for the spin chain including the irrelevant terms
with largest scaling dimensions is given by
\begin{eqnarray}
\label{B67}
&& H = \frac{v}{2} \int dx \l[\Pi^2+(\partial_x\phi)^2 -\frac{h}{\pi
    Rv}\partial_x\phi\r]+ \lambda_1 \int dx \cos\l(\frac{2\phi}{R}\r) \nn \\
&&  + \frac{\alpha}{\pi^2}\int dx \, \bigg[(\partial_x\phi_R)^2(\partial_x\phi_L)^2 +a(K)(\partial_x^2\phi_R)(\partial_x^2\phi_L)\bigg] \\
&& +\frac{\beta}{\pi^2}\int dx \bigg[(\partial_x\phi_R)^4+(\partial_x\phi_L)^4+b(K)((\partial_x^2\phi_R)^2+(\partial_x^2\phi_L)^2)\bigg] \nn
\end{eqnarray}
with a priori unknown amplitudes $\lambda_1,\alpha,\beta$. Here we have
separated the terms with integer scaling dimensions into a part where right-
and left-fields are mixed and a part where they remain separated. This will
allow us to determine the amplitudes in section \ref{Ampl} by using the fact
that no right and left mixing occurs at the free fermion point. The
coefficients $a(K),b(K)$ depend on the Luttinger parameter $K$ and describe
the relative weight between those terms in each bracket with different
symmetries. When we shift the $\phi$-field again as in (\ref{B28b}) the
magnetic field appears in the irrelevant terms.

We start with the umklapp term. First order perturbation theory in $\lambda_1$
leads to the correction
\begin{equation}
\label{B71}
F_1^{(1)} = \lambda_1 T\int_0^\beta\int_0^\infty d^2x \l\langle\cos\l(\frac{2\phi}{R}+\frac{2Khx}{v}\r)\r\rangle_0 
\end{equation}
where $\langle\cdots\rangle_0$ means the correlation function calculated for
the free boson model. This is the term which has already been considered in
\cite{FujimotoEggert} and \cite{FurusakiHikihara} to calculate the leading
contribution to the boundary susceptibility and specific heat for
$1/2<\Delta<1$ at zero magnetic field. We will derive here the more general
expression for the free energy contribution at finite temperature and finite
magnetic field. Using (\ref{B59}) we obtain at zero temperature
\begin{equation}
\label{B72}
E_1^{(1)} = \lambda_1\int_0^\infty dx
\frac{\cos\l(\frac{2Khx}{v}\r)}{(2x)^{2K}} \; .
\end{equation} 
To obtain the result at small finite temperatures we can use the standard
conformal mapping of the complex plane onto a cylinder to replace the
correlation function (\ref{B59}) by its finite temperature counterpart so that
(\ref{B72}) becomes
\begin{equation}
\label{B76}
F_1^{(1)} = \lambda_1\int_0^\infty dx
\frac{\cos\l(\frac{2Khx}{v}\r)}{\l[\frac{v}{\pi T}\sinh\l(\frac{2\pi T
    x}{v}\r)\r]^{2K}} \; .
\end{equation}
The integral is only convergent if $0<2K<1$. On the lattice convergence will
be insured by a lattice cutoff $\sim a$ as lower bound of integration. We can
then make the substitution $x=v\,\mbox{arcsinh}(u)/(2\pi T)$ and use partial
integration to separate the cutoff dependent and independent parts. The cutoff
independent part is expected to be the same for all $K$ and given by
\begin{eqnarray}
\label{B77}
\fl F_1^{(1,conv)} =
  \frac{\lambda_1\pi}{4}\frac{\Gamma(1-2K)\l[\csc(\pi
  K-\im\frac{Kh}{2T})+\csc(\pi K+\im
  \frac{Kh}{2T})\r]}{\Gamma\l(1-K-\im\frac{Kh}{2\pi
  T}\r)\Gamma\l(1-K+\im\frac{K h}{2\pi T}\r)}\,\l(\frac{2\pi
  T}{v}\r)^{2K-1} \; . 
\end{eqnarray}
There is no bulk contribution in this order of perturbation theory. Taking
derivatives with respect to $h$ and $T$, respectively, the known results for
the boundary susceptibility and specific heat are obtained
\cite{FujimotoEggert,FurusakiHikihara}.

To obtain a correction to the bulk free energy we have to go to second order
in $\lambda_1$. As the bulk part is not influenced by the boundary conditions
and we have already obtained the leading correction to the boundary part in
first order we will use PBC to calculate this correction. In this case we have
translational invariance leading to
\begin{eqnarray}
\label{B79}
f_1^{(2)} &=& -\frac{\lambda_1^2}{2}\int d^2x
\l\langle\cos\l(\frac{2\phi(x,\tau)}{R}+\frac{2Khx}{v}\r)\cos\l(\frac{2\phi(0,0)}{R}\r)\r\rangle_0
\nn \\
&=& -\frac{\lambda_1^2}{4}\int d^2x \cos\l(\frac{2Khx}{v}\r)
\l\langle\e^{2\rmi\phi(x,\tau)/R}\;\e^{-2\rmi\phi(0,0)/R}\r\rangle_0
\end{eqnarray}
We first consider the case $T=0$ where this correlation function can be
obtained from (\ref{B55}) and (\ref{B61}). The imaginary-time integral is
given by
\begin{equation}
\label{B80}
\int_0^\infty \frac{d\tau}{(x^2+v^2\tau^2)^{2K}} =
\frac{K\sqrt{\pi}\Gamma(2K-1/2)}{v\Gamma(2K+1)}\frac{1}{x^{4K-1}} \; .
\end{equation}  
This integral is always convergent because we assume $x\geq c$ where $c$ is a
lattice cutoff of order $a$. The cutoff independent part of (\ref{B79}) is then given
by
\begin{equation}
\label{B81}
\fl e_1^{(2,conv)} =
\frac{\lambda_1^2}{2}\frac{K\sqrt{\pi}\Gamma(2K-1/2)}{v\Gamma(2K+1)}\Gamma(2-4K)\cos(2K\pi)\l(\frac{2hK}{v}\r)^{4K-2}
\; .
\end{equation} 
For finite temperatures we can use again the conformal mapping of the plane
onto a cylinder for the correlation function. The free energy then becomes
\begin{equation}
\label{B83}
\fl f_1^{(2)} = -\frac{\lambda_1^2}{4}\l(\frac{\pi T}{v}\r)^{4K}\!\! \int_0^\infty
\!\int_0^\beta  \frac{\cos\frac{2Khx}{v}\, dxd\tau}{\l[\sinh\frac{\pi
    T}{v}(x+\im v\tau)\sinh\frac{\pi T}{v}(x-\im v \tau)\r]^{2K}}.
\end{equation}  
To evaluate this integral it is convenient to use the imaginary part of the
retarded instead of the imaginary time correlation function and to introduce
new variables $u_1=x-vt$, $u_2=-x-vt$ leading to 
\begin{eqnarray}
\label{B83.2}
f_1^{(2)} &=& -\frac{\lambda_1^2}{4v}\l(\frac{\pi T}{v}\r)^{4K}\sin(2\pi
    K) \\
&\times&\l[\int_0^\infty \frac{\exp(\im Khu/v)\, du}{\sinh^{2K}\frac{\pi
    T}{v}u}\r]\l[\int_0^\infty \frac{\exp(-\im Khu/v)\,
    du}{\sinh^{2K}\frac{\pi T}{v}u}\r] \; .\nn
\end{eqnarray}
This type of integral can be found to be given by \cite{Gradshteyn}
\begin{equation}
\label{B83.3}
\int_0^\infty du \frac{\exp(\im uz)}{\sinh^{2K}\frac{\pi
    T}{v}u}=\frac{2^{2K-1}v}{\pi T} B\l(K-\im\frac{vz}{2\pi T},1-2K\r)
\end{equation}
where $B(x,y)=\Gamma(x)\Gamma(y)/\Gamma(x+y)$. This allows us to obtain the
final result 
\begin{eqnarray}
\label{B83.4}
f_1^{(2)} &=& -\frac{\lambda_1^2}{4v}\l(\frac{2\pi T}{v}\r)^{4K-2}\sin(2\pi
K)\Gamma^2(1-2K) \\
&&\times \frac{\Gamma\l(K-\im\frac{Kh}{2\pi T}\r)\Gamma\l(K+\im\frac{Kh}{2\pi
    T}\r)}{\Gamma\l(1-K-\im\frac{Kh}{2\pi T}\r)\Gamma\l(1-K+\im\frac{Kh}{2\pi
    T}\r)} \; .\nn
\end{eqnarray}

The other irrelevant terms in (\ref{B67}) with amplitudes $\alpha,\beta$ can
be expressed as 
\begin{eqnarray}
\label{B110}
&& (\partial_x\phi_R)^4+(\partial_x\phi_L)^4 = \frac{1}{16}\l[2(\partial_{x} \phi)^4 -\frac{12}{v^2}(\partial_{x}
\phi)^2(\partial_\tau \phi)^2 +\frac{2}{v^4}(\partial_\tau \phi)^4 \r] \nn \\
&& (\partial_x\phi_R)^2(\partial_x\phi_L)^2 = \frac{1}{16}\l[(\partial_{x} \phi)^4 +\frac{2}{v^2}(\partial_{x}
\phi)^2(\partial_\tau \phi)^2 +\frac{1}{v^4}(\partial_\tau \phi)^4 \r] \; .
\end{eqnarray} 
When we now shift the boson field according to (\ref{B28b}), terms with an odd
number of $\phi$-fields appear. Such terms do not contribute when calculating
their expectation value in first order perturbation theory, because the free
boson Hamiltonian consists of an even number of $\phi$-fields. The operators
$(\partial_{x} \phi)^4, (\partial_{x} \phi)^2(\partial_\tau \phi)^2,
(\partial_\tau \phi)^4$ and
$(\partial_x^2\phi_R)^2,(\partial_x^2\phi_L)^2,(\partial_x^2\phi_R)(\partial_x^2\phi_L)$
independent of $h$ will only yield a $T^4$-contribution to the bulk part of
the free energy \cite{Lukyanov} and a $T^3$-contribution to the boundary part.
The $T^4$-term in the bulk free energy is sub-leading compared to the
$T^2$-term in (\ref{f3.2}) and we will ignore it. The same is true for the
$T^3$-term in the boundary free energy which will be sub-leading compared to a
$T^2$-term which cannot be determined within the field theory approach as
discussed at the end of subsection \ref{freeE}.

We therefore have to consider only the following three terms
\begin{equation}
\label{B84}
 \frac{AK^2h^4}{4\pi^2 v^4} \quad , \quad \frac{3AKh^2}{\pi v^2} (\partial_x \phi)^2
 \quad , \quad \frac{CKh^2}{2\pi v^2} (\partial_\tau \phi)^2
\end{equation}
where we have defined
\begin{equation}
\label{B111}
A=\frac{\alpha+2\beta}{16\pi^2} \quad , \quad
C=\frac{2\alpha-12\beta}{16\pi^2v^2} \; .
\end{equation}
The first term in (\ref{B84}) yields a temperature independent contribution
given by
\begin{equation}
\label{B85}
f_{A,a}^{(1)} = \frac{AK^2}{4\pi^2 v^4}h^4 \; .
\end{equation}
The next term gives
\begin{equation}
\label{B86}
f_{A,b}^{(1)} = \frac{AT}{L} \frac{3Kh^2}{\pi v^2}\int d^2 x \langle
(\partial_x\phi)^2\rangle \; .
\end{equation}
The correlation function $\langle (\partial_x\phi)^2\rangle$ at finite
temperature can be obtained by taking derivatives of the correlation function
in Eq.~(\ref{B52}) and applying the conformal mapping. To evaluate (\ref{B86})
we will do a point splitting by a small parameter $\epsilon$ in the
correlation function. This leads to 
\begin{eqnarray}
\label{B89}
f_{A,b}^{(1)} &=& -\frac{3KAh^2T^2}{2v^4L}\int_0^L dx \l(\frac{1}{\sinh^2\frac{\pi T}{v}\epsilon} +
\frac{1}{\sinh^2\frac{\pi T}{v}(2x+\epsilon)}\r) \; .
\end{eqnarray}
Doing the integral and then expanding in $\epsilon$ yields 
\begin{equation}
\label{B91}
f_{A,b}^{(1)} = -\frac{3AKh^2}{2\pi^2 v^2\epsilon^2}-\frac{3AKh^2}{4\pi^2
  v^2 L\epsilon} + \frac{AKh^2}{2v^4}T^2+\frac{1}{L}\frac{3AKh^2}{4\pi v^3}T+\mathcal{O}(\epsilon)
  \; .
\end{equation}
There are two different ways to interpret this result: One might think of
$\epsilon$ as a sort of lattice cutoff. Then one would say that the first term
yields a correction to the constant in the bulk susceptibility which depends
on this cutoff. However, we have already included all such corrections by
using the exact value for the Luttinger parameter $K$, so this term has to be
ignored. The second term shows that there is also a constant in the boundary
susceptibility, however, it is also cutoff dependent. This means that we cannot
determine this constant term within the field theory as expected from the
more general discussion in subsection \ref{freeE}. The last two terms which are
cutoff independent are therefore the only contributions we can really
determine within the field theory approach. Another point of view is that we
should use normal ordered operators. In this case we have to subtract the
$T=0$ correlations from (\ref{B89}) thus exactly cancelling the first two terms
in (\ref{B91}). In the limit $\eps\rightarrow 0$ only the two terms
independent of $\eps$ would then remain.

The calculation for the third term in (\ref{B84}) is analogous and the result
is
\begin{eqnarray}
\label{B94}
f_C^{(1)}&=&
\frac{CKh^2}{4\pi^2\eps^2}-\frac{CKh^2}{8\pi^2L\eps}-\frac{CKh^2T^2}{12v^2}+\frac{1}{L}\frac{CKh^2T}{8\pi
  v} \; .
\end{eqnarray}

With the help of (\ref{B111}) we can summarise our results for the bulk part
of the free energy as follows
\begin{eqnarray}
\label{B115}
\fl f_{\mbox{\tiny bulk}} = e_0-\frac{Kh^2}{4\pi
  v}+(\alpha+2\beta)\frac{K^2h^4}{64\pi^4 v^4} -\frac{\pi T^2}{6v}+(\alpha
  +6\beta)\frac{Kh^2T^2}{48\pi^2 v^4} \\
 \fl \qquad - \frac{\lambda_1^2}{4v}\frac{\sin(2\pi
K)\Gamma^2(1-2K)\Gamma\l(K-\im\frac{Kh}{2\pi T}\r)\Gamma\l(K+\im\frac{Kh}{2\pi
    T}\r)}{\Gamma\l(1-K-\im\frac{Kh}{2\pi T}\r)\Gamma\l(1-K+\im\frac{Kh}{2\pi
    T}\r)}\l(\frac{2\pi T}{v}\r)^{4K-2}  \nn
\end{eqnarray}
where $e_0$ is the ground state energy at zero magnetic field which is known
from BA. The boundary terms calculated by perturbation theory are obtained for the model on
the semi-infinite line, i.e., with one boundary. We therefore have to multiply
these results by 2. The boundary free energy is then given by
\begin{eqnarray}
\label{B118}
\fl F_B = E_0^B -\mathcal{B} h^2 +\alpha\frac{Kh^2T}{8\pi^3 v^3}+\mathcal{O}(T^2,T^4,h^2T^2,\cdots) \\
\fl \qquad + \frac{\lambda_1\pi}{2}\frac{\Gamma(1-2K)\l[\csc(\pi
  K-\im\frac{Kh}{2T})+\csc(\pi K+\im
  \frac{Kh}{2T})\r]}{\Gamma\l(1-K-\im\frac{Kh}{2\pi T}\r)\Gamma\l(1-K+\im\frac{K h}{2\pi T}\r)}\l(\frac{2\pi
  T}{v}\r)^{2K-1} \nn
\end{eqnarray}
where $E_0^B$ is the boundary ground state energy in zero magnetic field which
we will calculate by BA in the next section. The constant $\mathcal{B}$ cannot
be obtained within the field theory approach but will also be calculated in
the next section by BA. $\mathcal{O}(T^2,T^4,h^2T^2,\cdots)$ denotes the
regular terms which have been argued to be present in the boundary free energy
based on the results for the free fermion model (\ref{f0.2}). These terms
remain undetermined here. The amplitudes $\lambda_1,\alpha$ and $\beta$, on
the other hand, can be determined by comparing the field theory with the Bethe
ansatz result as has been first shown by Lukyanov \cite{Lukyanov} for the
periodic chain. We will show in section \ref{Ampl} that the amplitudes
obtained by comparing (\ref{B115}) and (\ref{B118}) with BA calculations for
the open chain are consistent with Lukyanov's result.
\section{The Bethe ansatz solution}
\label{BA}
In this section, leading and next-leading terms in a small-field expansion of
the ground-state energy are calculated within the Bethe ansatz solution for
the open chain. Using the Wiener-Hopf procedure, we obtain exact expressions
for both the bulk and the boundary contributions. Section \ref{nextlead}
contains an erratum to appendix B in \cite{BortzSirker}.

In order to introduce notations and to make this article self-contained, the
Bethe ansatz is shortly reviewed at the beginning, largely following
\cite{gau83}. We proceed to the thermodynamic limit afterwards. Leading and
next-leading contributions to the bulk and boundary susceptibility in terms of
the magnetic field are calculated subsequently.

\subsection{Coordinate Bethe ansatz for the open $XXZ$-chain}
The Hamiltonian \refeq{B1} conserves the total spin in $z$-direction $S^z=\sum_{j=1}^N S_j^z$, so that the eigenvalue of $S^z$ is a good quantum number. Let us denote an eigenstate with $S^z=N/2-M$ by $|M\rangle$. This eigenstate is a superposition of states with $M$-many $\down$-spins, which we denote by 
\be
|n_1,\ldots,n_M\rangle&=& S_{n_1}^-\ldots S_{n_M}^-|0\rangle\nn
\ee
where $|0\rangle$ is the fully polarised state with $M=0$. Thus
\be
|M\rangle = \sum_{\{n\}} a(n_1,\ldots,n_M) |n_1\ldots n_M\rangle = :  \sum_{\{n\}} a\{n\}|\{n\}\rangle.\nn
\ee
In terms of the coefficients $a\{n\}$, the eigenvalue equation $H|M\rangle = E|M\rangle$ reads
\be
\frac{J}{2} \sum_{\{n'\}}\l(a\{n'\}-\Delta a\{n\}\r) + \frac{J}{4} \Delta (N-1) a\{n\}&=& E a\{n\}\label{evp}.
\ee
Consider the simplified case 
\be
1<n_1<\ldots<n_M<N, \; n_{j+1}-n_{j}\geq 1\label{order}.
\ee
Then the $\{n'\}$ are given by the $\{n\}$, with $n_\al'=n_\al \pm 1$, so that 
\be
a\{n\}=\exp\l[ \rmi \sum_{j=1}^M k_j n_j \r]\label{5}
\ee
and 
\be
\fl\sum_{\al=1}^M a(n_1,\ldots,n_\al+1,\ldots,n_M) +a(n_1,\ldots,n_\al-1,\ldots,n_M)-2 \Delta a(n_1,\ldots,n_\al,\ldots,n_M)\nn\\
=2J E a(n_1,\ldots,n_M)\label{aeq}.
\ee
Inserting \refeq{5} yields
\be
E=J \sum_{j=1}^M \cos k_j + J\Delta\l(-M+\frac{N-1}{4}\r)\label{en}.
\ee
To solve the full eigenvalue problem \refeq{evp}, two complications arise:
\begin{itemize}
\item[i)] Flipped spins may be nearest neighbours. After acting with $H$ on a
  configuration with adjacent spins, amplitudes are obtained where the flipped
  spins are equal. These amplitudes must vanish. We thus first extend
  \refeq{order}, \be 1\leq n_1 \ldots\leq n_M \leq N.  \ee and then solve the
  eigenvalue problem by requiring that \refeq{aeq} and therefore \refeq{en}
  still hold. This means that the following terms in \refeq{aeq} must be zero:
  \be
  a(\ldots,n_\al+1,n_\al +1,\ldots)-\Delta a(\ldots,n_\al,n_\al+1,\ldots);\nn\\
  a(\ldots,n_\al,n_\al,\ldots)-\Delta a(\ldots,n_\al,n_\al+1,\ldots)\nn \ee
  which leads to 
\be \fl a(\ldots,n_\al+1,n_\al
  +1,\ldots)+a(\ldots,n_\al,n_\al,\ldots)-2\Delta
  a(\ldots,n_\al,n_\al+1,\ldots)=0.\label{8} 
\ee
\item[ii)] We have to deal with open boundaries. Therefore, we extend the lattice to include the sites $0$ and $N+1$, where transitions to these sites are excluded:
\be
a(0,n_2,\ldots) - \Delta a(1,n_2,\ldots) &=&0 \label{9first}\\
a(n_1,\ldots,n_{M-1},N+1)-\Delta a(n_1,\ldots,n_{M-1},N)&=& 0\label{9sec}
\ee
\end{itemize} 
We expect that i) leads to a scattering phase in the amplitudes whereas ii) yields the ``quantisation condition'' of the $k$s. 

In order to solve i), Bethe \cite{Bethe} made the ansatz
\be
a(n_1,\ldots,n_M)&=& \sum_{P\in \pi_M} A(P) \exp\l[ \rmi \sum_{\al=1}^M k_{P\al} n_\al\r]\label{ans},
\ee
where the sum carries over all permutations of $M$ integers.
Then \refeq{8} reads
\be
\fl\sum_P A(P)\l( \te^{\rmi (k_{P \al} + k_{P(\al+1)})} -2 \Delta \te^{\rmi k_{P(\al+1)}} +1\r) \te^{\rmi (k_{P1} n_1 + \ldots +  (k_{P\al} + k_{P(\al+1)})n_\al +\ldots)}=0\label{13},
\ee
where $n_\al=n_\al +1$. Since the terms $A(P)$ and $A(PP_{\al,\al+1})$ have the same $\{n\}$ dependence, \refeq{13} is fulfilled if
\be
\fl A(P) \l( \te^{\rmi(k_{P\al} + k_{P(\al+1)})} - 2 \Delta \te^{\rmi k_{P(\al+1)}} +1\r) \nn\\
\fl \qquad+ A(P P_{\al+1}) \l( \te^{\rmi(k_{P(\al+1)} + k_{P \al})} - 2 \Delta \te^{\rmi k_{P\al}} +1\r)=0\nn,
\ee
from which we conclude that
\be
A(P P_{\al+1}) = -A(P) \te^{-\rmi \theta_{P\al\,,P(\al+1)}}\label{app}
\ee
with the scattering phase
\be
\te^{-\rmi \theta_{P\al\,,P(\al+1)}}=\frac{\te^{\rmi(k_{P\al} + k_{P(\al+1)})} -2 \Delta \te^{\rmi k_{P(\al+1)}}+1}{\te^{\rmi(k_{P(\al+1)} + k_{P\al})} -2 \Delta \te^{\rmi k_{P\al}}+1}\label{15}.
\ee
{}From \refeq{app}, one concludes that 
\be
A(P)&=& \te^{\frac{\rmi}{2} \sum_{\al<\beta} \theta_{P\al,P\beta}}\label{14}.
\ee
If the chain was infinitely long, boundary conditions would not matter and one would choose the $k$s arbitrarily real (mod $2\pi$). However, to carry out the thermodynamic limit properly and to obtain the boundary contribution to the ground-state properties, one applies boundary conditions for the finite system. Here, these are open boundary conditions  (for periodic boundary conditions, cf. \cite{gau83}).

Before continuing, let us parametrise the $k_\al$ and $\theta_{\al,\beta}$ by roots $\lambda_\al$, such that \refeq{15} is fulfilled:
\be
\te^{\rmi k_\al} &=& \frac{\sinh(\la_\al+\rmi \g/2)}{\sinh(\la_\al-\rmi \g/2)} \label{18}\\
\te^{\rmi \theta(\la_\al,\la_\beta)}&=& \frac{\sinh\l(\la_\al - \la_\beta + \rmi \g\r)}{\sinh\l(\la_\al - \la_\beta - \rmi \g\r)}\label{19}
\ee

In order to meet the requirement ii), the ansatz \refeq{ans} is modified such that
\be
\fl a(n_1,n_2,\ldots,n_M) = \sum_{\{\eps\}} C(\eps_1,\ldots,\eps_M) \sum_P \te^{\frac{\rmi}{2} \sum_{\al<\beta} \theta(\la_\al-\la_\beta)+\rmi \sum_\al k_{P\al} n_\al}\label{ans2}
\ee
with $k_\al = \eps_\al |k_\al|$, $\la_\al = \eps_\al|\la_\al|$ and we have defined
the $2^N$ sets $\{ \eps\}$, $\eps_j=\pm 1$. Note that we have made use of
\refeq{14} in \refeq{ans}. It is important to note that due to the sum over signs, \refeq{ans2} makes
sense only for $k_\al\neq 0,\pm\pi$, that is $\la_\al\neq 0,\pm \rmi \pi/2$ for all $\al$.  

The modified ansatz \refeq{ans2} does not influence \refeq{18}, \refeq{19}. However, it allows to meet the open boundary conditions \refeq{9first}, \refeq{9sec}. Eq. \refeq{9first} yields
\be
\fl C(\eps_1,\ldots,\eps_{P_1},\ldots,\eps_M) \l( \sum_{\al\neq1} \te^{\frac{\rmi}{2} \theta(\la_{P1}-\la_{P\al})}-\Delta \te^{\rmi k_{P1}}\r) \nn\\
+ C(\eps_1,\ldots,-\eps_{P_1},\ldots,\eps_M) \l( \sum_{\al\neq1} \te^{-\frac{\rmi}{2} \theta(\la_{P1}+\la_{P\al})}-\Delta \te^{-\rmi k_{P1}}\r) =0\label{aux}.
\ee
{}From \refeq{15},
\be
-\te^{-\rmi \theta(2\la_\al)} &=& \frac{1-\Delta \te^{-\rmi k_\al}}{1-\Delta \te^{\rmi k_\al}}\label{21},
\ee
such that \refeq{aux} becomes
\be
\frac{C(\eps_1,\ldots,-\eps_{P_1},\ldots,\eps_M)}{C(\eps_1,\ldots,\eps_{P_1},\ldots,\eps_M)}&=& \te^{\frac{\rmi}{2} \sum_{\beta \neq \al} (\theta(\la_\al - \la_\beta) + \theta(\la_\al + \la_\beta) + 2 \theta(2\la_\al))}\nn
\ee
with the unique solution
\be
C(\eps_1,\ldots,\eps_M)&=& \te^{-\frac{\rmi}{2} \sum_{\al\leq \beta} \theta(\la_\al + \la_\beta)}\nn.
\ee
We thus have the amplitude
\be
\fl a\{n\} &=& \sum_P \sum_{\{\eps\}} \te^{\frac{\rmi}{2} \sum_{\al<\beta}( \theta(\la_\al - \la_\beta) - \theta(\la_\al + \la_\beta))-\frac{\rmi}{2} \sum_\al \theta(2\la_\al) +\rmi \sum_\al n_\al k_{P\al}}\nn.
\ee
This expression is substituted into Eq.~\refeq{9sec} to obtain
\be
\fl\te^{\rmi N k_{PM}}\l(\te^{\rmi k_{PM}}-\Delta\r) \te^{\frac{\rmi}{2} \sum_{\al\leq M} \theta(\la_{P\al}-\la_{PM})-\theta(\la_{P\al}+\la_{PM})} \nn\\
+ \te^{-\rmi N k_{PM}}\l( \te^{-\rmi k_{PM}} - \Delta\r) \te^{\frac{\rmi}{2} \sum_{\al\leq M} \theta(\la_{P\al}+\la_{PM})-\theta(\la_{P\al}-\la_{PM})}&=& 0\label{25}.
\ee
Note that the $k$s are parametrised by the $\la$s via \refeq{18}. Using \refeq{21}, Eq.~\refeq{25} is written as
\be 
\frac{a^{2(N+1)}(\la_k,\g/2)}{a^{2(N+1)}(\la_k,-\g/2)}\,\frac{a(2\la_k,- \g)}{a(2\la_k, \g)} = -\frac{q_M(\la_k+\rmi\g) q_M(-\la_k- \rmi\g)}{q_M(\la_k-\rmi\g)
  q_M(-\la_k+\rmi \g)}\label{cbae}, \ee
with the definitions 
\be \fl
 a(\la,\mu):=\sinh(\la+\rmi \mu)\quad
, \quad q_M(\la):= \prod_{j=1}^M \sinh(\la-\la_j)\nn\,.  
\ee
This equation is equivalent to 
\be 
\fl\l[\frac{a(\la_k, \g/2)}{a(\la_k,-\g/2)}\r]^{2 N}\,\frac{a(2\la_k,\g)
  a(\la_k,\pi/2- \g/2)
  a(\la_k,\pi/2-\g/2)}{a(2\la_k,-\g) a(\la_k,-\pi/2+\g/2)a(\la_k,-\pi/2+\g/2)}\nn\\
\fl\qquad = -\frac{q_M(\la_k+\rmi\g) q_M(-\la_k-\rmi \g)}{q_M(\la_k-\rmi\g)
  q_M(-\la_k+\rmi \g)}\label{bae1}, 
\ee 
which also follows from the algebraic Bethe ansatz \cite{Sklyanin}. The Eqs. \refeq{cbae} are the above mentioned ``quantisation conditions'' on the $\la$s and therefore on the $k$s. 

The energy \refeq{en}, including a magnetic field $h$ along the $S^z$-direction, reads in terms of the $\la$s:
\be E&=&J\l[-\sum_{j=1}^M\frac{\sin^2
  \gamma}{\cosh(2 \lambda_j)-\cos\gamma}
+\frac{N-1}{4}\,\cos\gamma\r]-h S^z\label{ej}\\
S^z&=&N/2-M\label{sz}.  
\ee
The distribution of the roots has been investigated in \cite{BortzSirker}. There it has been shown that the equations \refeq{bae1} can be symmetrised by introducing the set of roots $\{v_1,\ldots,v_N\}:=\{-\la_{N/2},\ldots,-\la_{1},\la_1,\ldots,\la_{N/2}\}$,
whose elements are distributed symmetrically on the real axis w.r.t.~the
origin. The $v_j$ are the $N$ real solutions
to the equations 
\be \fl \l[\frac{a(\la_k, \g/2)}{a(\la_k,-\g/2)}\r]^{2 N}\,\l[\frac{a(v_m,\pi/2-\g/2)\,a(v_m,\g/2)}{a(v_m,-\pi/2+\g/2)\,a(v_m,-\g/2)}\r]=\frac{q_N(v_m+\rmi\g)}{q_N(v_m-\rmi\g)}\label{bae2}.
\ee
Note that $\pm\rmi \pi/2, 0$ are solutions of \refeq{bae2}. However, these solutions are not permitted within the
Bethe ansatz as explained after Eq.~\refeq{ans2}.\footnote{Within the algebraic Bethe ansatz, the creation
operators which are used to create eigenvectors by acting on a
reference state are identically zero at spectral parameters $0,\pm \rmi \pi/2$. This can be seen directly from
Sklyanin's work \cite{Sklyanin}.} This is a direct consequence of the broken
translational in the open system.
\subsection{$T=0$ properties in the thermodynamic limit}
To pass to the thermodynamic limit, it is convenient to define the density of
roots on the real axis, $\rho(x)$. In \cite{BortzSirker}, the following linear integral equation has been derived for $\rho(x)$:
\be 
\vartheta(x,\g)+\frac{1}{2N}\l[\vartheta(x,\g)+\vartheta(x,\pi-\g)+\vartheta(x,2\g)\r]\nn\\
\qquad=\rho(x)+\int_{-B}^B\vartheta(x-y,2\g)\rho(y)\,\d
y\label{int1}, 
\ee 
where 
\be 2\pi \rmi \,\vartheta(x,\g)&:=&\frac{2\rmi
  \sin\g}{\cosh 2x-\cos\g}=\frac{\d}{\d x}\ln \frac{\sinh(x+\rmi
  \g/2)}{\sinh(x-\rmi\g/2)}=\rmi\frac{\d}{\d x} k(x).  
\ee
This equation is valid including the order $\Or(1/N)$, that is, including the bulk and the boundary contributions. 

Equation \refeq{int1} is a linear integral
equation with two unknowns, $B$ and $\rho$.  In a first step, \refeq{int1} is
solved for $B=\infty$; in a second step, $\rho(x)$ is obtained depending on
the parameter $B$ and the dependence of $B$ on the magnetic field $h$ is
calculated. We will see that $B=\infty$ corresponds to $h=0$, and a finite
magnetic field $h>0$ induces a finite $B<\infty$. Finally, the susceptibility
$\chi(h)$ is deduced. This procedure is reviewed in \cite{tak99}.

It will be shown later that instead of dealing with $\rho$, all
quantities we are interested in can be expressed more conveniently by
$g_+(x):=\theta(x)\rho(x+B)$. The calculation of these functions is done by
Fourier transformation, 
\be 
\rho(x)&=&\frac{1}{2\pi} \i \widetilde
\rho(k)\te^{-\rmi kx}\,\d k\nn\; .  
\ee 
Let us first consider the case
$B=\infty$. It is straightforward to solve
\refeq{int1} in Fourier space, where 
\be 
\widetilde
\vartheta(k,\gamma)&=&\frac{\sinh(\pi/2-\g/2)k}{\sinh\pi k/2} \nn.  
\ee 
We
denote the solution of \refeq{int1} for $B=\infty$ by $\rho_0$ and find 
\be
\wt \rho_0(k)=\wt s(k)+\frac{1}{2N} \,\frac{\cosh\g k/4
  \,\cosh(\pi/4-\g/2)k}{\cosh\g k/2\,\cosh(\pi-\g)k/4}\label{rho0}, 
\ee 
with
\be \wt s(k):=\frac{1}{2\cosh\g k/2},\;
s(x)=\frac{1}{2 \g \cosh\pi x/\g}\nn.
\ee 
Note that $\i\rho_0(x)\d x=1/2+1/(2N)$, which corresponds to $M=N/2$ in Eq.~\refeq{bae2}. Putting this into Eq.~\refeq{sz}, one has $S^z=0$. Thus a vanishing magnetic field corresponds to $B=\infty$. This means that in order to perform a small field expansion $|h|\ll \al$ (where $\al$ is some scale which is determined later), one has to expand asymptotically the energy at $B\to \infty$. 

To do so, we now consider the case $B<\infty$, i.e., a finite magnetic field. Using
\refeq{rho0} we can rewrite \refeq{int1} as \be
\rho(x)&=&\rho_0(x)+\int_{|y|>B}\kappa(x-y)\rho(y)\,\d y\label{lie}\\
\kappa(x)&:=&\frac{1}{2\pi} \i \frac{\sinh(\pi/2-\g)k}{2\cosh \g
  k/2\,\sinh(\pi-\g)k/2}\,\te^{-\rmi kx}\,\d k\label{defkap}.  \ee We now
introduce the functions  
\be
\rho(x+B)=:g(x)\equiv g_+(x)+g_-(x)\nn\\
g_+(x)=\theta(x) g(x)\,,\qquad g_-(x)=\theta(-x) g(x)\label{defgpm}\,.
\ee
Then $g(x)$ satisfies the equation 
\be 
\fl
g(x)=\rho_0(x+B)+\ip \kappa(x-y)g_+(y)\,\d y+\ip \kappa(x+y+2B)g_+(y)\,\d
y\label{geq}.  
\ee 
The driving term $\rho_0(x+B)$ can be expanded in powers of
$\exp\l[B\r]$. Since \refeq{geq} is linear in $g$ and $\rho_0$, we make the
ansatz $g=g^{(1)}+g^{(2)}+\ldots$, where superscripts denote increasing powers
of $\exp\l[B\r]$. Then 
\be
\fl g^{(1)}(x)=\l[\rho_0(x+B)\r]^{(1)}+\ip\kappa(x-y)g_+^{(1)}(y)\,\d y\label{g1}\\
\fl g^{(n)}(x)= \l[\rho_0(x+B)\r]^{(n)}+\ip \kappa(x+y+2B)g_+^{(n-1)}(y)\,\d
y\nn\\
+\ip \kappa(x-y)g_+^{(n)}(y)\,\d y\label{gng}.  
\ee 
Thus in each order, a linear
integral equation of Wiener-Hopf-type has to be solved. This technique is
explained for example in \cite{krei62,roo69}. 

The first two orders of $\wt g_+(k)$ read:
\be
\wt g_+^{(1)}(k)&=&G_+(k)\l[\wt \rho_0(k)G_-(k)\te^{-\rmi k
    B}\r]_+^{(1)}\label{g1f}\\
\wt g_+^{(2)}(k)&=&G_+(k)\l\{\l[\wt \rho_0(k)\,G_-(k)\,\te^{-\rmi k
  B}\r]_+^{(2)}\r.\nn\\
& &\l.+\l[\wt \kappa(k)\,\wt g^{(1)}(-k)\,G_-(k)\,\te^{-2\rmi
  kB}\r]_+^{(2)}\r\}\,\label{g2}, 
\ee
where the indices $\pm$ are defined by
\be
f_\pm(k):=\pm\frac{\rmi}{2\pi} \i \frac{f(q)}{k-q\pm\rmi \epsilon}\,\d
q\label{pint}. 
\ee
The functions $G_\pm(k)$ are obtained from the factorisation $1-\wt \kappa=1/(G_+ G_-)$, they read
\be
G_+(k)&=&\frac{\sqrt{2(\pi-\g)}\,\G(1-\rmi k/2)}{\G(1/2-\rmi \g
  k/(2\pi))\,\G(1-\rmi k(\pi-\g)/(2\pi))}\,\te^{-\rmi a k}\label{gp}\\
a&=& \frac12\l[\frac\g\pi\ln(\pi/\g-1)-\ln(1-\g/\pi)\r]\nn\\
G_-(k)&=& G_+(-k)\label{gmm}.
\ee 
We restrict ourselves to the calculation of $\wt g_+$, which is
sufficient for our purposes.

The bracket $\l[\ldots\r]_+^{(1)}$ in \refeq{g1f} and the first term in \refeq{g2} is evaluated using
\refeq{pint}. We thus have to find the residues of 
\be
\frac{1}{2\cosh \frac{\g k}{2}},\; \frac{\cosh \frac{\g k}{4} \, \cosh \l(\frac\pi4-\frac\g2\r)k}{\cosh\frac{\g k}{2} \, \cosh\frac{\pi-\g}{4} k}
\ee
at the poles closest to the real axis in the lower half plane (the first term in the above line accounts for the bulk contribution, the second for the boundary contribution from \refeq{rho0}). This is done straightforwardly for the bulk part: the poles are located at 
\be
k_n^{(1)}=- \rmi (2 n+1)\pi/\g,\;n=0,1,\ldots\label{k1}.
\ee
The relevant poles of the boundary part, however, depend on whether $\g<\pi/3$ or $\g>\pi/3$: Poles at $k_n^{(1)}$ are found as well as poles at 
\be
k_m^{(2)}=-\rmi 2(2m+1)\pi/(\pi-\g),\; m=0,1,\ldots.\label{pole2}
\ee
For $\g>\pi/3$, the leading pole is $k_0^{(1)}$, whereas for $\g<\pi/3$, the
pole $k_0^{(2)}$ is leading. Finally double poles occur at \be \frac{\g}{\pi}
&=& \frac{2 n+1}{4m +2 n+3},\; k_{m,n}^{(1,2)}= -\rmi (4 m + 2 n +3),\;
m,n=0,1,\ldots \label{doub}, \ee where the leading and next-leading ones occur
for $\g=\pi/3,3 \pi/5$. In the ongoing, we will concentrate on the single
poles, i.e. $\g\neq \pi/3$ in the leading order and $\g\neq \pi/3,3\pi/5$ in
the next-leading order. The extension of the results to the cases $\g=\pi/3,
\,3\pi/5$ is discussed in section \ref{main}. We first consider the
leading-order contributions \refeq{g1f}, before proceeding to the next-leading
corrections \refeq{g2}. The isotropic case $\g=0$ is treated separately.
\subsubsection{The leading orders}
\label{leadord}
By taking only the poles $k_0^{(1)}$, $k_0^{(2)}$ nearest to the real axis into
account, one finds for $\g\neq \pi/3$: 
\be 
\fl \wt g_+^{(1)}(k)=G_+(k)\l\{\displaystyle\frac{a_0}{k+\rmi \pi/\g}\te^{-\pi/\g
    B}+\frac{1}{2N}\l[\frac{a_1}{k+\rmi\pi/\g}\te^{-\pi B/\g}\r.\r.\nn\\
\fl \qquad \l.\l.+\frac{b_1}{k+\rmi
    2\pi/(\pi-\g)}\te^{-2\pi B/(\pi-\g)}\r]\r\} \label{gp1}.
\ee 
We could proceed analogously for $\g=\pi/3$ by evaluating the residue at the double pole \cite{BortzSirker}. However, it turns out to be more convenient to include the point $\g=\pi/3$ only at the very end, once the susceptibility has been calculated. The constants in \refeq{gp1} are given by 
\numparts
\be
a_0&=&\frac{\rmi}{\g}G_-(-\rmi \pi/\g)\label{c1}\\
a_1&=& \frac{\sqrt{2} \rmi}{\g} G_-(-\rmi
\pi/\g)\frac{\sin\pi^2/(4\g)}{\cos(\pi^2/(4\g)-\pi/4)}\label{c2}\\
b_1&=& \frac{2\rmi}{\pi-\g}\tan\pi\g/(\pi-\g)\,G_-(-\rmi 2\pi/(\pi-\g))\label{c3}.
\ee
\endnumparts
We can now compute $s^z:=S^z/N$ and $e:=E/N$ from
\refeq{ej},\refeq{sz}: 
\be
\fl s^z=1/2-\int_{-B}^B\rho(x)\d x+1/(2N)\label{sz2}\\
\fl e=-h s^z-\frac{J\sin\g}{2}\int_{-B}^B \vartheta(x,\g)\rho(x)\d
x+\frac{J}{4}\l(\cos\g+\frac{2-\cos \g}{N}\r)\label{gs1}.  
\ee 
We insert
\refeq{lie} into \refeq{sz2} to obtain 
\be 
s^z=\frac{\pi}{\pi-\g}\, \wt
g_+(0)\label{sz3},  
\ee 
which is an exact statement, including all orders $\wt g^{(n)}$. 
It is convenient to calculate $e-e_0$, where
$e_0:=e(h=0)$ is the ground state energy at zero magnetic field. From \refeq{rho0}, \refeq{gs1} one finds
\be
\fl e_0= \frac{J}{4}\l( \cos \g + \frac{2-\cos \g}{N}\r) -\frac{J \sin \g}{2} \i \vartheta(x,\g)\, \rho_0(x)\, \d x\nn\\
\fl\qquad = \frac{J}{4} \cos\g - \frac{J \sin \g}{4\pi} \i\frac{\sinh\l(\pi/2 - \g/2\r)k}{2\cosh \g k/2 \, \sinh \pi k/2} \d k+ \frac{1}{N}\l[ \frac{J(2-\cos \g)}{4} \r.\nn\\
\fl\qquad \l.  - \frac{J \sin \g}{4 \pi} \i\frac{\cosh \g k/4\, \cosh(\pi/4 - \g/2) k}{2\cosh \g k/2 \,\cosh(\pi-\g) k/4}\,\frac{\sinh\l(\pi/2 - \g/2\r)k}{\sinh \pi k/2} \d k\r]\label{gsdet}
\ee
We use again \refeq{lie} which yields 
\be 
\fl e-e_0=-h
s^z+\frac{4J\pi\sin\g}{\g}\ip \frac{g_+(x)}{\cosh(x+B)\pi/\g}\,\d
x\nn\\
\fl \qquad =-\frac{h\pi}{\pi-\g} \l(\wt g^{(1)}_+(0)+\wt g^{(2)}_+(0)\r) + \frac{8\pi J \sin\g}{\g}\l[\l(
\wt
g^{(1)}_+(\rmi \pi/\g)+\wt g^{(2)}(\rmi \pi/\g) \r)\te^{-\pi B/\g}\r.\nn\\
\fl \qquad\qquad -\l.\wt g_+^{(1)}(3 \rmi \pi/\g) \te^{-3\pi B/\g}
+\Or\l(\te^{-3\pi B/\g}\wt g^{(2)}\r)\r]\label{gse}, 
\ee 
where in the last
equation we restrict ourselves to the given orders. Now $B$ is treated as a
variational parameter and is determined in such a way that \be
\frac{\partial}{\partial B}(e-e_0)=0\label{var}.  \ee In this section we
consider only the leading order in \refeq{gse}. Inserting \refeq{gp1},
\refeq{sz3}, \refeq{gse} in \refeq{var}, $B$ is obtained as a function of $h$,
\be
B&=&-\frac\g\pi \ln\frac h\al\label{bh1} \\
\al&:=& \frac{2\pi J \sin\g}{\g} \frac{(\pi-\g)}{\pi} \frac{G_+(\rmi
  \pi/\g)}{G_+(0)}\; . \label{alh} \ee Thus $\alpha$ sets the scale for $h$ (this scale $\al$ should not be confused with the amplitude $\alpha$ introduced in section \ref{Bos}). The
restriction to the leading orders in $\exp[-B]$ is equivalent to the leading
orders in $h$ in the limit $|h|\ll \al$.

One now makes use of \refeq{alh} to determine $s^z(h)$ from \refeq{sz3}, and
therefrom $\chi(h)=\partial s^z/\partial h$. Inserting the expressions for $G_\pm$ from Eqs.~\refeq{gp}, \refeq{gmm} we find 
\numparts
\be
\fl\chi_{\mbox{\footnotesize bulk}}=\frac{\g}{(\pi-\g)\pi J
  \sin\g} \; .\label{chib}
\ee
This result is well known, see, for example, \cite{tak99}.
The boundary contribution is given by
\be
\fl\chi_B(h)= \frac{\g}{J(\pi-\g)\pi \sqrt{2}\,
  \sin\g}\,\frac{\sin\pi^2/(4\g)}{\cos(\pi^2/(4\g)-\pi/4)}\nn\\
\fl \qquad \qquad+\frac{2\g\sqrt{\pi}}{(\pi-\g)^2}\tan\frac{\pi\g}{\pi-\g}\frac{1}{\bar\al}\,\frac{\G\l(\pi/(\pi-\g)\r)}{\G\l(1/2+\g/(\pi-\g)\r)}\,(h/\bar\al)^{-(\pi-3\g)/(\pi-\g)}
\label{chi1} 
\ee 
for $\g\neq \pi/3$ with 
\be \fl \bar\al= 2J (\pi-\g)\sqrt\pi\,
\frac{\sin \g}{\g}\, \frac{\G(1+\pi/(2\g))}{\G(1/2+\pi/(2\g))} \label{al}. 
\ee
\endnumparts
Note that the first term in
\refeq{chi1}, which is independent of the magnetic field $h$, is the leading
contribution for $\g>\pi/3$ (pole closest to the real axis in \refeq{g1f}). For
$\g<\pi/3$ the second term dominates. In how far the result \refeq{chi1} also yields the next-leading corrections for the whole range of $0<
\g<\pi$ cannot be answered at this point: so far, we have only focused on the leading poles. Next-leading corrections will be discussed in the next section. 

The constant contribution in \refeq{chi1}, as well as the pre-factor of the $h$-dependent part, show divergences for certain values of $\g$. We will comment on these in section \ref{main}.

\subsubsection{The next-leading orders} 
\label{nextlead}
In this section, the Wiener-Hopf result \refeq{g2} in \refeq{gse}, \refeq{sz3}
is used to calculate next-leading corrections to \refeq{chi1}. This section
also constitutes an erratum to appendix B in \cite{BortzSirker}.

Let us start with the bulk contribution. We already commented on the calculation of the first term in \refeq{g2} after that equation. The second term involves the residues of $\wt \kappa(k)$ at the poles closest to the real axis. As can be seen from \refeq{defkap}, poles occur at $k_n^{(1)}$ (cf. Eq.~\refeq{k1}) and at 
\be
k_l^{(3)}=-\rmi \frac{2\pi}{\pi-\g} l, \; l=1,2,\ldots\label{kl3}.
\ee
This situation is very similar to the leading order of the boundary contribution. 

The boundary contribution is treated analogously. The only difference is that the function $\wt g_+^{(1)}(k)$ given in \refeq{gp1} contains two terms in $1/N$. The first term obviously yields exponents in analogy to the bulk. The second one has poles given in Eq.~\refeq{pole2} which have to be combined with those in Eq.~\refeq{kl3}. 

In the ongoing, we will focus on the amplitudes of the most important next-leading terms, that is, we consider the residues due to $k_0^{(1)}$ and $k_0^{(3)}$ in the second term in \refeq{g2}. Calculating the residua mentioned above and putting everything together, one ends up with
\be
\fl \wt g_+^{(2)}(k)=G_+(k)\nn\\
\fl\qquad\times\l\{\l(\frac{a_{0,1}}{k+\rmi 3\pi/\g}+ \frac{a_{0,2}}{k+\rmi \pi/\g}\r)\te^{-3\pi
  B/\g}+\frac{a_{0,3}}{k+\rmi
  2\pi/(\pi-\g)}\,\te^{-(\pi/\g+4\pi/(\pi-\g))B}\r.\nn\\
\fl\qquad  +\frac{1}{2N}\l[\l(\frac{a_{1,1}}{k+\rmi
    3\pi/\g}+\frac{a_{1,2}}{k+\rmi\pi/\g}\r)\te^{-3\pi
    B/\g}+\frac{a_{1,3}}{k+\rmi
    2\pi/(\pi-\g)}\,\te^{-(\pi/\g+4\pi/(\pi-\g))B}\r.\nn\\
\fl\qquad+\l(\frac{b_{1,1}}{k+\rmi 6\pi/(\pi-\g)}\r.\nn\\
\fl\qquad\l.\l.\l.+\frac{b_{1,3}}{k+\rmi2\pi/(\pi-\g)}\r)\te^{-6\pi/(\pi-\g)B}
+\frac{b_{1,2}}{k+\rmi\pi/\g}\te^{-2(\pi/\g+\pi/(\pi-\g))B}\r]
\r\}\label{g2f}\\
\fl a_{0,1}=-\frac{\rmi}{\g} \,G_-\l(-\rmi \frac{3\pi}{\g}\r)\\
\fl a_{0,2}=
\frac{\rmi}{2\g\pi}\,\tan\frac{\pi^2}{2\g}\,G_-^3\l(-\rmi\frac\pi\g\r)\\
\fl a_{0,3}=
\frac{\rmi}{\pi(\pi+\g)}\,\tan\frac{\pi\g}{\pi-\g}\,G_-\l(-\rmi\frac\pi\g\r)\,G^2_-\l(-\rmi\frac{2\pi}{\pi-\g}\r)\\
\fl a_{1,1}= \rmi\frac{2\,\sin\pi/4\,\sin
  3\pi^2/(4\g)}{\g\,\cos(3\pi^2/(4\g)+\pi/4)}\,G_-\l(-\rmi
\frac{3\pi}{\g}\r)\\
\fl a_{1,2}= \frac{a_1}{2\pi} \tan\frac{\pi^2}{2\g}
G^2_-\l(-\rmi\frac\pi\g\r)\\
\fl a_{1,3}=
\frac{a_1\g}{\pi(\pi+\g)}\,\tan\frac{\pi\g}{\pi-\g}\,G_-^2\l(-\rmi\frac{2\pi}{\pi-\g}\r)\\
\fl b_{1,1}=\rmi\frac{2}{\pi-\g}\,\tan\frac{3\g\pi}{\pi-\g}\,G_-\l(-\rmi\frac{6\pi}{\pi-\g}\r)\\
\fl b_{1,2}=\frac{b_1(\pi-\g)}{\pi(\pi+\g)}\,\tan\frac{\pi^2}{2\g}\,G_-^2\l(-\rmi\frac\pi\g\r)\\
\fl b_{1,3}= \frac{b_1}{4\pi} \,\tan\frac{\pi\g}{\pi-\g}\,G_-^2\l(-\rmi\frac{2\pi}{\pi-\g}\r)\label{lasteq},
\ee
This expression for $\wt g_2$ is inserted into \refeq{gse}, where we now have
to keep all the indicated terms.
Then $B$ as a function of $h$ is derived. In section \ref{leadord}, we found that
this relationship is the same both for the boundary and for the bulk in the
leading order. This is no
longer true when next-leading terms are considered. For the bulk we obtain
\be
\fl\al\te^{-\pi B/\g}=
h\l(1+A_1\l(\frac{h}{\al}\r)^2+A_2\l(\frac{h}{\al}\r)^{4\g/(\pi-\g)}\r)\label{hb1}\\
\fl A_1= \frac{a_{0,2}}{a_0}+\frac{G_-(-\rmi 3\pi/\g)}{G_-(-\rmi\pi/\g)}\,,\;A_2= \frac{\pi-\g}{2\g}\,\frac{a_{0,3}}{a_0}\nn.
\ee
Care has to be taken by inverting the relation $h=h(B)$ in order to get $B=B(h)$ for the boundary: this is done by performing a large-$B$-expansion in $h=h(B)$. The next-leading terms in this expansion depend on the range of $\g$, whether $\g<\pi/3$ or $\g>\pi/3$. In the final result $B=B(h)$, the parameter $\g$ must enter uniquely. Then we find
\be
\fl \al\te^{-\pi B/\g}=
h\l(1+A_1\l(\frac{h}{\al}\r)^2+A_2\l(\frac{h}{\al}\r)^{4\g/(\pi-\g)}\r)\label{hb2}\\
\fl A_1=\frac{a_{1,2}}{a_1}+\frac{G_-(-\rmi 3\pi/\g)}{G_-(-\rmi\pi/\g)}\,,\;A_2= \frac{\pi-\g}{2\g}\,\frac{a_{1,3}}{a_1}.
\ee
Here terms of higher order, previously given in \cite{BortzSirker}, have been discarded for the reason mentioned above.

Combining these equations with \refeq{sz3}, one finds
\be
\fl s_{\mbox{\footnotesize bulk}}^z(h)= \sqrt{\frac{2}{\pi(\pi-\g)}}\l\{G_-(-\rmi \pi/\g)\frac h\al \r.\nn\\
\fl\qquad+ \l(\frac{1}{\pi}\tan\frac{\pi^2}{2\g}\,G_-^3\l(-\rmi \frac\pi\g\r)+\frac23G_-\l(-\rmi \frac{3\pi}{\g}\r)\r)\l(\frac h\al\r)^3\nn\\
\fl\qquad +\frac{\pi-\g}{\pi(\pi+\g)}\,\tan\frac{\pi\g}{\pi-\g}\,G_-\l(-\rmi\frac\pi\g\r)\,G_-^2\l(-\rmi\frac{2\pi}{\pi-\g}\r)\,\l(\frac h\al\r)^{1+4\g/(\pi-\g)}\label{sbz}\\
\fl s^z_B(h)= -\rmi\sqrt{\frac{1}{2\pi(\pi-\g)}}\l\{\g a_1\frac h\al+\frac{\pi-\g}{2}\,b_1\,\l(\frac h\al\r)^{2\g/(\pi-\g)}\r.\nn\\
\fl\qquad+\l(2\g a_{1,2}+\frac{\g}{3}\,a_{1,1}+\g a_1\frac{G_+(\rmi 3\pi/\g)}{G_+(\rmi \pi/\g)}\r)\l(\frac h\al\r)^3\nn\\
\fl \l.\qquad+\Or\l(\l(\frac{h}{\al}\r)^{1+\frac{4\g}{\pi-\g}},\l(\frac{h}{\al}\r)^{\frac{6\g}{\pi-\g}},\l(\frac{h}{\al}\r)^{2+\frac{2\g}{\pi-\g}}\r)\r\}\label{sbbz}
\ee
These results correct those of \cite{BortzSirker}. From these expressions, $\chi_{\mbox{\footnotesize bulk}}$ and $\chi_B$ can be obtained:
\be
\fl \chi_{\mbox{\footnotesize bulk}}(h)= \sqrt{\frac{2}{\pi(\pi-\g)}}\l\{G_-(-\rmi \pi/\g)\frac 1\al \r.\nn\\
\fl\qquad+ \frac{3}{\al}\l(\frac{1}{\pi}\tan\frac{\pi^2}{2\g}\,G_-^3\l(-\rmi \frac\pi\g\r)+\frac23G_-\l(-\rmi \frac{3\pi}{\g}\r)\r)\l(\frac h\al\r)^2\nn\\
\fl\qquad +\frac{3\g+\pi}{\al\pi(\pi+\g)}\,\tan\frac{\pi\g}{\pi-\g}\,G_-\l(-\rmi\frac\pi\g\r)\,G_-^2\l(-\rmi\frac{2\pi}{\pi-\g}\r)\,\l(\frac h\al\r)^{4\g/(\pi-\g)}\label{chibz}\\\fl \chi_B(h)= -\rmi\sqrt{\frac{1}{2\pi(\pi-\g)}}\l\{\frac{\g a_1}{\al}+\frac{\g\,b_1}{\al}\,\l(\frac h\al\r)^{2\g/(\pi-\g)-1}\r.\nn\\
\fl\qquad+\frac{3}{\al}\l(2\g a_{1,2}+\frac{\g}{3}\,a_{1,1}+\g a_1\frac{G_+(\rmi 3\pi/\g)}{G_+(\rmi \pi/\g)}\r)\l(\frac h\al\r)^2\nn\\
\fl \l.\qquad+\Or\l(\l(\frac{h}{\al}\r)^{\frac{4\g}{\pi-\g}},\l(\frac{h}{\al}\r)^{\frac{6\g}{\pi-\g}-1},\l(\frac{h}{\al}\r)^{1+\frac{2\g}{\pi-\g}}\r)\r\}\label{chibbz}
\ee

\subsubsection{Isotropic $XXX$-point}
\label{islim}
The isotropic case $\g=0$ (i.e. $\Delta=1$) is treated in the same manner as the
anisotropic case $\g\neq 0$. Therefore, we rescale \refeq{ej} by $\la_j\to\g\la_j$. This is equivalent to
substituting $k\to k/\g$ in Fourier space. Then 
\be
\wt s(k)&=&\frac{1}{2\cosh k/2}\nn\\
\wt \rho_0(k)&=& \wt s(k)+\frac{1}{2N}\,\frac{1}{2\cosh k/2}
\l(1+\te^{-|k|/2}\r)\label{isrho0}\; .  
\ee 
Whereas the analyticity properties
of the bulk contribution to \refeq{isrho0} are qualitatively the same as in
\refeq{rho0}, the boundary contribution shows, besides poles, a cut along the
imaginary axis. The functions $g_+^{(1,2)}$ are determined as in the anisotropic case, where now the integrals in \refeq{g1f}, \refeq{g2} encircle the cuts of \refeq{isrho0} and of 
\be
\kappa(x)=\frac{1}{2\pi} \i \frac{\te^{-|k|/2}}{2 \cosh k/2} \te^{-\rmi k x}\d k
\ee
(that is the isotropic limit of \refeq{defkap}). Accordingly, the functions $G_\pm$ from Eq.~\refeq{gp}, \refeq{gmm} now read
\be
G_+(k)&=& \sqrt{2\pi} \frac{(-\rmi k)^{-\rmi k/(2\pi)}}{\G(1/2+\rmi
  k/(2\pi))}\,\te^{-\rmi a k}\nn\\
a&=& -\frac{1}{2\pi}-\frac{\ln(2\pi)}{2\pi}\nn\\
G_-(k)&=&G_+(-k)\nn
\ee
Let us first consider the bulk contribution. We find
\be
\fl \wt g_+^{(1)}(k) + \wt g_+^{(2)}(k)= G_+(k) \l( \frac{a_0}{k+\rmi \pi} \te^{-\pi B} - \frac{\rmi G_-(-3\rmi \pi)}{k+3\rmi \pi} \te^{-3\pi B}\r) \nn\\
\fl \qquad \qquad \qquad \qquad  +\l\{ \begin{array}{ll}
  \frac{\beta_0 \te^{-\pi B}}{k B^2}, & k\neq 0 \\
  \frac{\beta_1 \te^{-\pi B}}{B} + \beta_2 \frac{\ln B}{B^2} \te^{-\pi B} + \beta_3 \frac{\te^{-\pi B}}{B^2},&k=0 
  \end{array}\r.\label{g12}
\ee
with
\be
\beta_0 &=& \frac{a_0}{16\pi^2} G_-^2(0)\qquad \beta_1 = \frac{a_0}{4\rmi \pi^2}\nn\\
\beta_2 &=& -\frac{a_0}{8\rmi \pi^3} \qquad \beta_3 = \frac{a_0}{8\rmi \pi^3}(-\ln\pi+1)\nn.
\ee
Inserting \refeq{g12} into \refeq{gse} and performing the maximisation condition \refeq{var}, one obtains
\be
\te^{-\pi B}&=& \frac{h}{\al}\l( 1+\frac{\al_1}{\ln h/\al} + \al_2 \frac{\ln|\ln h/\al|}{\ln^2 h/\al} + \frac{\al_3}{\ln^2 h/\al}\r)\label{bhis}
\ee
with
\be
\al_1 &=& -\rmi \frac{\beta_1 \pi^2}{a_0}= -\frac14\nn\\
\al_2 &=& \rmi \frac{\beta_2 \pi^3}{a_0} = -\frac18\nn\\
\al_3 &=& \frac{\pi^2}{a_0}\l( \rmi \beta_3 \pi+ \rmi \beta_1 - 2\beta_0 - \rmi \pi \beta_2 \ln \pi\r)=-\frac{1}{8}\nn
\ee
Combining \refeq{g12}, \refeq{bhis} and \refeq{gse}, one obtains for the bulk susceptibility
\be
\fl \chi_{{\rm bulk}}=\chi_{{\rm bulk}}(h=0) \l( 1+\frac{2\al_1}{\ln h/\al} + 2\al_2 \frac{\ln |\ln h/\al|}{\ln^2 h/\al} + \frac{\al_4}{\ln^2 h/\al}\r)\label{chibis},
\ee
with $\al_4=5/16$, $\chi_{{\rm bulk}}(h=0)=J/\pi^2$, and $\al=\sqrt{2 \pi^3/\te}$. We now set $\al= h_0 \delta$, where $\delta$ is determined such that the term $\sim \ln^{-2} h/\al$ in \refeq{chibis} is absorbed into the $1/\ln( h/h_0)$-term. This prescription fixes the scale uniquely. One finds $\delta=\exp(-5/8)$ and 
\be
 h_0 = \sqrt{2\pi^3} \te^{-9/8}.
\ee
Then Eq.~\refeq{chibis} reads
\be
 \chi_{{\rm bulk}}=\frac{1}{J\pi^2} \l( 1-\frac{1}{2\ln h/h_0} - \frac{\ln |\ln h/h_0|}{4\ln^2 h/h_0} \r)\nn.
\ee
Let us now consider the boundary contribution. 
In \refeq{g1f}, the $\l[\ldots\r]_+$-bracket yields
contributions $\Or(\exp\l[-const.\,B\r])$ from the poles, and algebraic
contributions due to the cut. The exponential contributions are clearly
sub-leading in comparison to the algebraic ones, so only the latter are
calculated in the following. Using the expressions for $G_{\pm}(k)$, we find (omitting the bulk
contribution) 
\be 
\fl\wt g_+^{(1)}(k)=\l\{\begin{array}{ll}
  G_+(0)(\al_1/B+\al_2(\ln B)/B^2+\al_3/B^2)& \\[0.2cm]
  +\Or((\ln B)/B^3,1/B^3),& k=0\\[0.2cm]
  \rmi \al_1 G_+(k)/(k B^2),& k\neq 0
\end{array}
\r.\nn\\
\fl\al_1= \frac{1}{\sqrt 2 \pi}\,,\;\al_2= -\frac{\sqrt2}{4\pi^2}\,,\;\al_3= \frac{1}{\sqrt 2\pi^2}\l(\ln2-\frac{1}{2}\ln(2\pi)\r)\,.\nn
\ee
Leading and next-leading contributions are here already contained in $\wt g_+^{(1)}$, so that we do not consider further corrections stemming from $\wt g_+^{(1)}$. From \refeq{sz3}, \refeq{gse}, \refeq{var} we obtain
\be
B&=&-\frac1\pi \ln\frac h\al\label{bh2}\\
\al^{-1}&=& \frac{G_+(0)}{2\pi J G_+(\rmi \pi)}\; . \label{alh2}
\ee
These equations are obtained by those from the anisotropic case, \refeq{bh1},
\refeq{alh}, by scaling $B\to\g B$ and sending $\g\to 0$ afterwards. Carrying out the same steps which lead to
\refeq{chi1}, one finds the boundary contribution 
\be
\fl s^z_B(h)=-\frac14\l(\frac{1}{\ln h/\wt h_0}+\frac{\ln|\ln h/\wt h_0|}{2\ln
  ^2h/\wt h_0}\r)+\mbox{o}(\ln^{-2}h) \label{szis}\\
\fl\chi_B(h)= \frac{1}{4h}\l(\frac{1}{ \ln ^2 h/\wt h_0}+\frac{\ln|\ln
  h/\wt h_0|}{\ln^3h/\wt h_0}-\frac{1}{2\ln^3h/\wt h_0}\r)+\mbox{o}\l(\frac{1}{h\ln^3h}\r)\label{chiis}\\
\fl \wt h_0=\al/\sqrt{2}=J\pi \sqrt{\pi/\te}\; . \label{h0}
\ee
The scale $\wt h_0$ has been chosen such that in \refeq{szis}, no terms
$\Or(\ln^{-2}h)$ occur. Note that $\wt h_0 \neq h_0$. The results \refeq{chiis}, \refeq{h0} agree with the TBA-work by Frahm et
al.~\cite{frah97} for $T=0$. Furthermore, agreement is found with \cite{FujimotoEggert,FurusakiHikihara,AsakawaSuzuki96b}, where scales which differ from ours \refeq{h0} by a constant factor were used.      
\section{Determining the amplitudes by comparing field theory and Bethe ansatz}
\label{Ampl}
Apart from the amplitudes $\lambda_1,\alpha,\beta$, also the bulk and boundary
part of the ground state energy as well as the constant term in the boundary
susceptibility have been left unknown in our final field-theory result
(\ref{B115}) and (\ref{B118}). These quantities have been calculated in the
last section by Bethe ansatz. In terms of the Luttinger parameter $K$ we can
write the ground state energy as 
\begin{equation}
\label{GZE}
e_0^{\mbox{\tiny bulk}} = -\frac{J}{4}\cos\frac{\pi}{K}-\frac{J}{8\pi}\int_{-\infty}^\infty
dk\frac{\sin\frac{\pi}{K}\sinh\frac{\pi}{2K}}{\sinh\frac{\pi
    k}{2}\cosh\frac{(K-1)\pi k}{2K}}
\end{equation}
and 
\begin{eqnarray}
\label{GZEB}
&&E_0^B = \frac{J}{4}\l(2+\cos\frac{\pi}{K}\r) \\
&&-\frac{J}{8\pi}\sin\frac{\pi}{K}\sinh\frac{\pi}{2K}\int_{-\infty}^\infty
dk\frac{\cosh\frac{(K-2)\pi k}{4K}\cosh\frac{(K-1)\pi k}{4K}}{\sinh\frac{\pi
    k}{2}\cosh\frac{\pi k}{4 K}\cosh\frac{(K-1)\pi k}{2K}} \; .\nn
\end{eqnarray}
The constant $\mathcal{B}$ in Eq.~(\ref{B118}) can also be expressed in term
of the Luttinger parameter
\begin{equation}
\label{B107}
\mathcal{B} = \frac{K}{2\pi v}\frac{1}{2\sqrt{2}}\frac{\sin\frac{\pi K}{4K-4}}{\cos\frac{\pi}{4K-4}} \; . 
\end{equation}

Next we determine the amplitudes. By taking the limit $T\rightarrow 0$ of Eq.~(\ref{B77}) we obtain
\begin{equation}
\label{B74}
E_1^{(1,conv)} = -\lambda_1
K^{2K}\Gamma(-2K)\sin(K\pi)\l(\frac{h}{v}\r)^{2K-1} \;.
\end{equation}
The amplitude $\lambda_1$ of the umklapp term can now be found by comparing
this with the BA result in Eq.~(\ref{chi1}). This leads to
\begin{equation}
\label{B108}
\lambda_1
=\frac{K\Gamma(K)\sin\pi/K}{\pi\Gamma(2-K)}\l[\frac{\Gamma\l(1+\frac{1}{2K-2}\r)}{2\sqrt{\pi}\Gamma\l(1+\frac{K}{2K-2}\r)}\r]^{2K-2}
\end{equation} 
and agrees with Lukyanov's result in Eq.~(2.24) of Ref.~\cite{Lukyanov} where
our amplitude $\lambda_1$ is related to his amplitude $\lambda$ by
$\lambda_1=v\lambda/2\pi$. Here we have determined $\lambda_1$ by comparing
with a boundary quantity. We can, of course, also compare the zero temperature
limit of the bulk contribution (\ref{B83.4}) with (\ref{sbz}). This yields the
same result as it should be.

The amplitudes of the terms with scaling dimension $4$ represent a more subtle
problem. From the zero temperature Bethe ansatz solution (\ref{sbz}) we can
only obtain the amplitude of the $h^4$-term in the free energy. According to
(\ref{B115}), however, this amplitude depends on $\alpha$ and $\beta$. One
possibility would be to calculate the amplitude of the $h^2T^2$-term in
$f_{\mbox{\tiny bulk}}$ from a finite temperature BA solution. Although such a
solution exists, an analytic formula for this term is difficult to obtain. We
therefore follow Lukyanov's idea to obtain both amplitudes from the $h^4$-term
in $f_{\mbox{\tiny bulk}}$ alone, by using symmetry arguments. The coefficient
of the $h^3$-contribution in (\ref{sbz}) consists of two terms. The first of
these terms vanishes at the free fermion point due to the
$\tan$-function.\footnote{This does not come as a surprise: The term has been
  obtained from the second bracket in (\ref{g2}). However, at the free
  fermion point $\g=\pi/2$, the integration kernel $\kappa$ vanishes
  identically, $\kappa|_{\gamma=\pi/2}\equiv 0$.} We can identify this term
as the one determining the amplitude $\alpha$, because $\alpha$ is associated
with a mixing of left- and right-movers and therefore has to vanish at the
free fermion point. The other term then determines $\beta$ leading to
\begin{equation}
\label{B112}
\alpha = -\pi^2v\tan\frac{\pi K}{2K-2} 
\end{equation}
and 
\begin{equation}
\label{B114}
\beta=-\frac{v\pi^2}{6K}\frac{\Gamma\l(\frac{3K}{2K-2}\r)\Gamma^3\l(\frac{1}{2K-2}\r)}{\Gamma\l(\frac{3}{2K-2}\r)\Gamma^3\l(\frac{K}{2K-2}\r)}
\; .
\end{equation}
These results also agree with Lukyanov's formula (2.24) in \cite{Lukyanov}
with $\alpha = -2\pi^3v\lambda_+$ and $\beta=-2\pi^3v\lambda_-$.

\section{Divergent amplitudes and logarithmic corrections}
\label{main}
{}From (\ref{B115}) and (\ref{B118}) the bulk and boundary susceptibilities for
zero magnetic field can be derived by taking derivatives with respect to $h$
leading to
\begin{eqnarray}
\label{B116}
\fl \chi_{\mbox{\tiny bulk}}(T,h=0) = \frac{K}{2\pi
  v}-(\alpha+6\beta)\frac{K T^2}{24\pi^2 v^4} +\frac{\lambda_1^2}{32\pi v^3}\Gamma^2(1/2-K)\nn \\
\fl \qquad \times\Gamma^2(1+K)\sin(2\pi K)\l[\Psi'(1-K)-\Psi'(K)\r]\l(\frac{\pi T}{v}\r)^{4K-4} 
\end{eqnarray}
and
\begin{eqnarray}
\label{B119}
\fl \chi_B(T,h=0) = 2\mathcal{B}-\alpha\frac{KT}{4\pi^3 v^3} + \lambda_1\frac{K\Gamma(K)\Gamma(3-2K)(\pi^2-2\Psi'(K))}{2v^2(2-1/K)\Gamma(2-K)}\left(\frac{2\pi
    T}{v}\right)^{2K-3} \; . 
\end{eqnarray}
Interestingly, the constant term $2\mathcal{B}$ is closely related to the
constant in the bulk susceptibility (see (\ref{B107})). The two terms become
equivalent at the free fermion point as expected from the calculations in
subsection \ref{freeE}. 
On the other hand, for $T=0$, bulk and boundary susceptibilities are given in Eqs.~\refeq{chibz}, \refeq{chibbz}. 

For certain anisotropies, terms in $\chi(T=0,h)$ and $\chi(T, h=0)$ show
divergences. In this section, we will show that these divergences cancel and
give rise to logarithmic corrections. In the first part, we will focus on
$\chi(T=0,h)$, whereas in the second part, $\chi(T, h=0)$ is treated. In order
to keep contact with the notations introduced in sections \ref{Bos} and
\ref{BA}, we will write our results in terms of $\g=\arccos \Delta$ for the
$T=0$ case, whereas $K=\pi/(\pi-\g)$ will be employed for $h=0$ at finite $T$.

\subsection{Logarithmic corrections in $\chi(T=0,h)$}
Let us come back to the enumeration of poles encountered within the
Wiener-Hopf procedure in the leading and next-leading orders,
Eqs.~(\ref{k1},\ref{pole2},\ref{kl3}). Combining the relation $B=B(h)$,
Eq.~\refeq{bh1}, with the functional dependence of the energy on $\wt g_+$ ,
Eq.~\refeq{gse}, we can read off the exponents of $h$-dependent contributions.
Namely, the poles \refeq{k1} lead to terms \be \sim h^{2n},\;
n=0,1,\ldots\label{con1} \ee 
both in the boundary and the bulk susceptibility. Furthermore, the $k_m^{(2)}$
from \refeq{pole2} lead to contributions 
\be 
\sim h^{\frac{\g(4m+3)
    -\pi}{\pi-\g}},\; m=0,1,\ldots\label{con2}  
\ee in the boundary part. As
far as the next-leading order is concerned, one determines the exponents to
the bulk susceptibility due to the second term in \refeq{g2} from the
combination of poles \refeq{k1}, \refeq{kl3}. This results in terms \be \sim
h^{2(n+2m+1)} + h^{2(n+2 \frac{\g}{\pi-\g}l)},\; n,m=0,1,\ldots\;; \;
l=1,\ldots\label{con3}.  \ee The next-leading order of the boundary
susceptibility shows, apart from terms similar to \refeq{con3}, additional
contributions of the form \be \fl \sim h^{4 n +1 +\frac{2\g}{\pi-\g}(2 m +1)}
+ h^{\frac{2\g}{\pi-\g}\l(2m+1 + 2l \r) -1},\;
n,m=0,1,\ldots\;l=1,\ldots\label{con4}.  \ee which result form combining
\refeq{pole2} with \refeq{k1} and \refeq{kl3}. Note that the pre-factors of the
leading terms in Eqs.~(\ref{con1},\ref{con2},\ref{con3},\ref{con4}) are those
which have been determined in section \ref{BA}. By carrying out the
Wiener-Hopf procedure to higher orders, it would be possible to obtain further
pre-factors, at the expense of more and more cumbersome calculations.

The enumeration of possible exponents in (\ref{con1},\ref{con2},\ref{con3},\ref{con4}) is far from complete: These exponents have been found from extrapolating the first and second order results of the perturbation expansion of $\wt g_+$, while using the first-order result \refeq{bh1} and the leading orders in \refeq{gse}. In higher orders of the $\wt g_+$-expansion, these exponents mix with each other and further orders appear.

So far, we excluded the double poles in Eq.~\refeq{doub}, which occur when two poles of the types (\ref{k1},\ref{pole2}) coincide, that is, when exponents stemming from (\ref{con1},\ref{con2}) cross. Again combining Eqs.~\refeq{bh1}, \refeq{gse} we obtain logarithmic contributions 
\be
\sim h^{2 n} \ln h \label{hlog},\; n=0,1,\ldots
\ee
to the boundary susceptibility. The same happens if exponents in the next-leading order contributions cross, therefore yielding logarithmic contributions of the form \refeq{hlog}, with $n\geq 1$ there, for both the bulk and the boundary.   
In order to obtain the coefficient of the logarithmic terms, it suffices to know the amplitudes of the terms whose exponents cross: They show divergences at the crossover points, which cancel to yield the logarithms. 

Consider the constant term in $\chi_B(h)$, cf. Eq.~\refeq{chibbz}. It displays poles at
\be
\g_n=\frac{\pi}{4 n+3}\label{gn}
\ee
with an accumulation point at $\g=0$, whereas the coefficient of the $h$-dependent contribution with exponent $2\g/(\pi-\g)-1$ has poles at 
\be
\g_m=\frac{2 m+1}{2 m+3}\pi\label{gm}
\ee
with an accumulation point at $\g=\pi$. Obviously, $\g_n=\g_m$ for $n=0=m$, which means $\g_0=\pi/3$ (the case $n=m=-1$, $\g_{-1}=\pi$, is excluded here). 
By setting $\g=\pi/3+\eps$ and expanding in $\eps$, the two divergences cancel and one obtains
\be
\chi_B(h)|_{\g=\pi/3}&=& -\frac{1}{\sqrt 3 \pi^2} \ln \l( \frac{h \te^{3-\pi/2}}{4\pi 3^{3/2}}\r)\label{chi11}.
\ee
As far as the other poles \refeq{gn}, \refeq{gm} are concerned, note that they coincide with \refeq{doub} for $m=0$ or $n=0$ there, respectively. This means that we expect higher-order terms, whose coefficients show poles at \refeq{gn}, such that logarithms of the form \refeq{hlog} occur, with $n\geq 1$ there.

A similar crossover at $\g=\pi/3$ happens in the bulk part in the next-leading order, when the exponents $2,4\g/(\pi-\g)$ cross in Eq.~\refeq{chibz}. Again divergences in the two corresponding amplitudes cancel, as can be seen by expanding $\g=\frac{\pi}{3}+\eps$. Then one finds
\be
\chi_{\mbox{\footnotesize bulk}}|_{\g=\pi/3}(h)&=& \frac{1}{\sqrt 3\pi} - \frac{2}{3 \sqrt 3 \pi^3} h^2 \ln \frac{h\te^{-1/3}}{4  \pi3^{3/2}}.
\ee

Finally, let us consider the additional crossover in the next-leading order of $\chi_B(h)$. In Eq.~\refeq{chibbz}, three exponents have been identified: $0,\,2\g/(\pi-\g)-1,\,2$. Contrary to the bulk case, there are now two crossover points: $\g=\pi/3$ and $\g=3\pi/5$. At $\g=\pi/3$, the (leading) logarithmic contribution has been identified in Eq.~\refeq{chi11}. At $\g=3\pi/5$, a (next-leading) logarithmic contribution appears, which is
\be
\label{K5.2}
\fl \chi_B(h)= \chi_0\l(1+h^2\l(\frac{135}{64} \frac{(3+\sqrt 5)\G^3(4/3)}{\pi^{3/2}(3+\sqrt 3)(5+2\sqrt 5)\G^3(11/6)}\r.\r.\nn\\
\fl\qquad \times \l(-18 \ln h -31+39 \ln 2+9 \ln\frac{(5+\sqrt 5)\pi\G^2(11/6)}{\G^2(1/3)} \r)\nn\\
\fl\qquad  +\l.\l.\frac{405}{64} \frac{(2+\sqrt 3)(3+\sqrt 5)\G^3(4/3)}{(3+\sqrt 3)\pi^{1/2} (5+2\sqrt 5) \G^3(11/6)} - \frac{27}{10} \frac{(1+\sqrt 3)(3+\sqrt 5)}{\pi (3+\sqrt 3)(5+2 \sqrt 5)}\r)\r)
\ee
with 
\be
\chi_0&=& \frac{40}{\pi} \frac{3+\sqrt 3}{\sqrt 2}(5+2\sqrt 5)(5+\sqrt{5})^{-7/2}.
\ee
This completes our discussion of the leading and next-leading terms in the susceptibilities on the basis of the Bethe-ansatz solution at $T=0$ for the anisotropic $XXZ$-case. 

\subsection{Logarithmic corrections in $\chi(T,h=0)$}
Let us start our discussion for finite temperatures with the bulk
susceptibility. The $T^2$-term in $\chi_{\mbox{\tiny bulk}}$ contains the
amplitude $\alpha$ which is defined in (\ref{B112}).  $\alpha$ is divergent
for $K=(2n+1)/2n$ with $n=1,2,\cdots$.\footnote{The amplitude $\beta$ does not
  show any divergences.} Obviously $\chi_{\mbox{\tiny bulk}}$ does not
diverge at these points so this divergence has to be cancelled by other
divergent terms. In fact, the $T^{4K-4}$-term has also divergences at
$K=(2m+1)/2$ with $m=1,2,\cdots$ due to the $\Gamma^2(1/2-K)$-factor. The only
point where both terms simultaneously diverge is therefore $K=3/2$ and we will
show that these divergences indeed cancel each other here and lead to a
logarithmic correction. But what happens to the other divergences? Let's
first consider the divergences in the $T^{4K-4}$-term. At $K=(2m+1)/2$ the
temperature dependence becomes $T^{4m-2}=T^2,T^6,T^{10},\cdots$. Terms like
$h^2T^6,h^2T^{10},\cdots$ appear in the free energy in perturbation theory in
the irrelevant terms with integer scaling dimension. For example, in $n$-th
order perturbation theory in the $(\partial_x\phi)^4$ operator we will get a
$h^2T^{4n-2}$-term.  These higher order corrections will interfere with the
$T^{4K-4}$-term and cancel the divergences at $K=(2m+1)/2$. The result will
be a $T^{4m-2}\ln T$ behaviour at these points. As the $T^2$-term dominates
for $K>2$ these are just logarithmic corrections in next leading
contributions.

Much more interesting are the divergences at $K=(2n+1)/2n$ in the $T^2$-term.
They yield $T^2\ln T$-terms and the points where this happens become dense for
$K\rightarrow 1$ ($\Delta\rightarrow 1$). The divergence at $K=3/2$ in the
$T^2$-term is cancelled by the divergence in the $T^{4K-4}$-term. In general,
the divergence at $K=(2n+1)/2n$ is cancelled by a $T^{4nK-4n}$-term which one
obtains in $2n$-th order perturbation theory in the $\cos(2\phi/R)$-operator.

The only case where we can give an explicit result for the logarithmic term is
$K=3/2$. Here we find
 \begin{eqnarray}
\label{B121}
\chi_{\mbox{\tiny bulk}}(T,h=0) &=&
\frac{1}{\sqrt{3}\pi}+\frac{142+24\gamma+12\ln(3888)-21\zeta(3)}{243\sqrt{3}\pi}T^2  \\
&&-\frac{8}{81\sqrt{3}\pi}T^2\ln T \; ,  \nn
\end{eqnarray} 
where $\gamma$ is Euler's constant and $\zeta(x)$ Riemann's zeta function.

In the boundary susceptibility similar things happen. Here even the constant
term (\ref{B107}) as determined by BA shows divergences at
$K=(4n-1)/(4n-2)=3/2,7/6,\cdots$ due to the $\cos$-term in the denominator, see also Eq.~\refeq{gn}.
The $T^{2K-3}$-term, on the other hand, has a divergent amplitude for
$K=(2n+1)/2$. Again all these divergences have to cancel each other leading
to logarithmic corrections. For $K=3/2$ we can derive an explicit result
 \begin{eqnarray}
\label{B122}
\fl \chi_B(T,h=0) =
\frac{-6+4\gamma+2\pi+2\ln(3888)-7\zeta(3)}{4\sqrt{3}\pi^2}- \frac{\ln T}{\sqrt{3}\pi^2} +\mathcal{O}(T\ln T)\; .
\end{eqnarray} 
Here the $T\ln T$-term stems from the term linear in $T$ whose amplitude also
diverges at $K=3/2$.

\begin{figure}[!htp]
\begin{center}
  \includegraphics*[width=0.75\columnwidth]{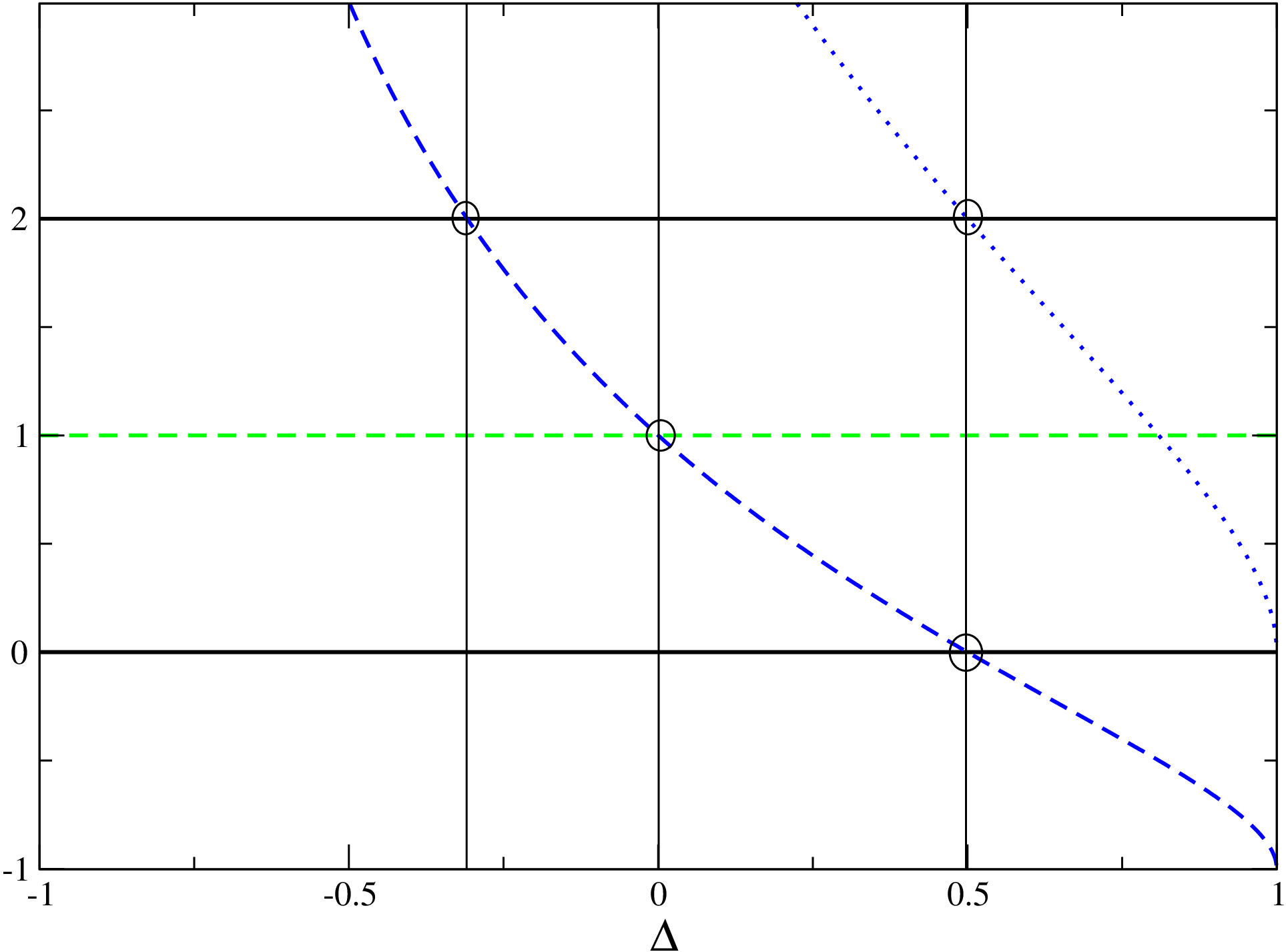} 
\caption{Crossover of critical exponents, depending on the anisotropy
  $\Delta$, in the leading orders. Black lines are those exponents which occur
  both in the bulk and boundary susceptibilities (namely 0 and 2). The leading
  non-integer exponent in the boundary contribution is dashed blue ($2 K-3$),
  the one in the bulk contribution dotted blue ($4 K-4$). Green is the
  exponent 1, which only occurs at finite $T$ in the boundary contribution.
  All crossover points are denoted by circles. The crossover between 1 and
  $2K-3$ at $\Delta=0$ has no consequences, since the corresponding amplitudes
  vanish. At all other crossover points, logarithmic corrections appear. The
  vertical lines denote $\Delta=(1-\sqrt{5})/4$
  ($\gamma=3\pi/5$), $\Delta=0$ and $\Delta=0.5$.}  
\end{center}
\label{exp}
\end{figure}

We summarise our results in Fig.~\ref{exp}.
\section{Comparison with numerical results}
\label{numerics}
\begin{figure}[!htp]
\begin{center}
\includegraphics*[width=0.99\columnwidth]{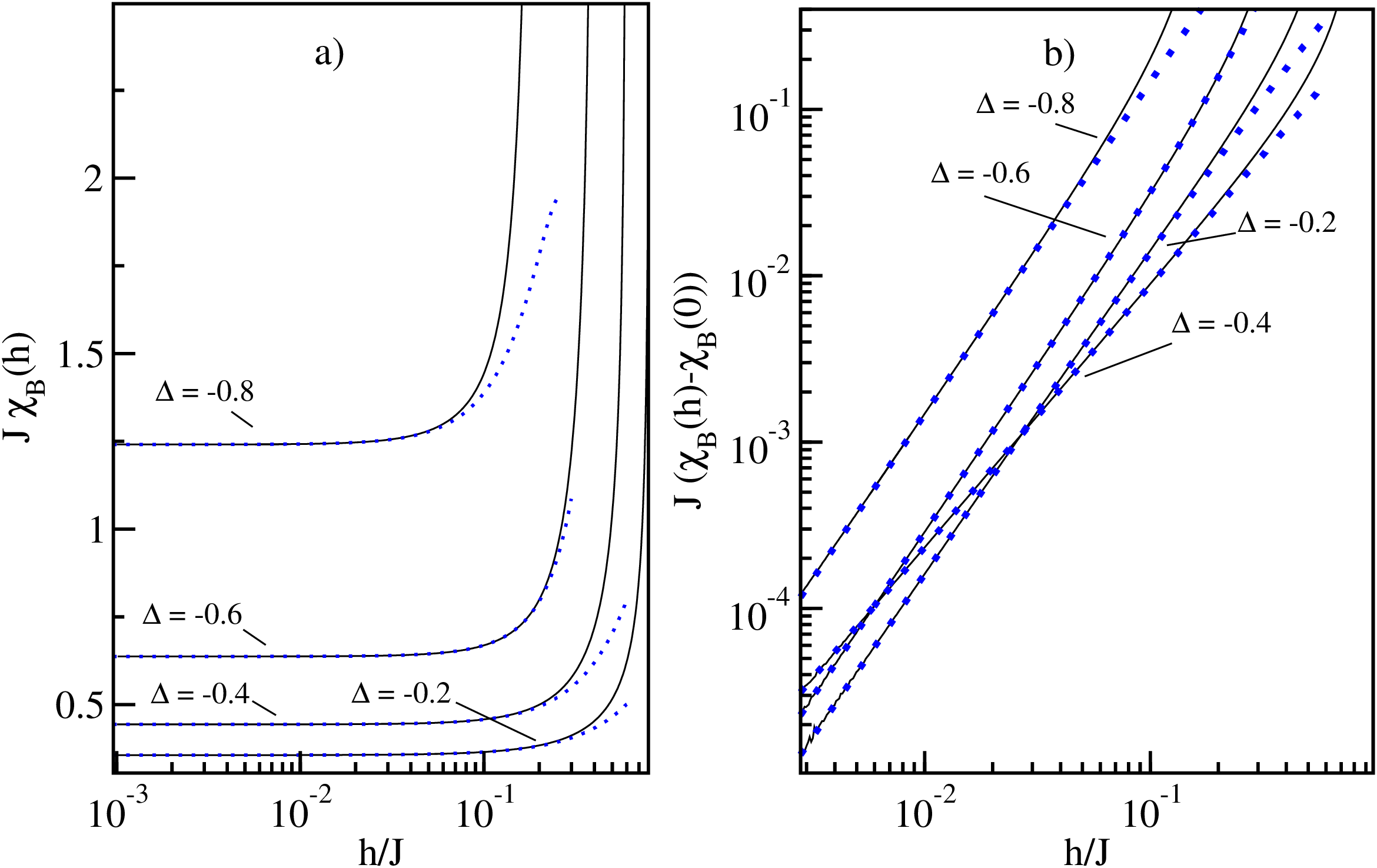}
\end{center}
\caption{a) Boundary susceptibility $\chi_B(h)$: Comparison between a numerical solution
  of the BA equations (black solid lines) and the analytical solution for
  $h/J\ll \alpha$ in the attractive regime (blue dotted lines). b) The
  difference $\chi_B(h)-\chi_B(0)$ on a double logarithmic scale. For
  $\Delta<\cos (3\pi/5)\approx -0.31$, the leading $h$-term has constant
  exponent 2, whereas for $\Delta>\cos (3\pi/5)$, the exponent depends on
  $\g$, according to Eq.~\refeq{chibbz}.}
\label{FigT0_1}
\end{figure}

\begin{figure}[!htp]
\begin{center}
\includegraphics*[width=0.75\columnwidth]{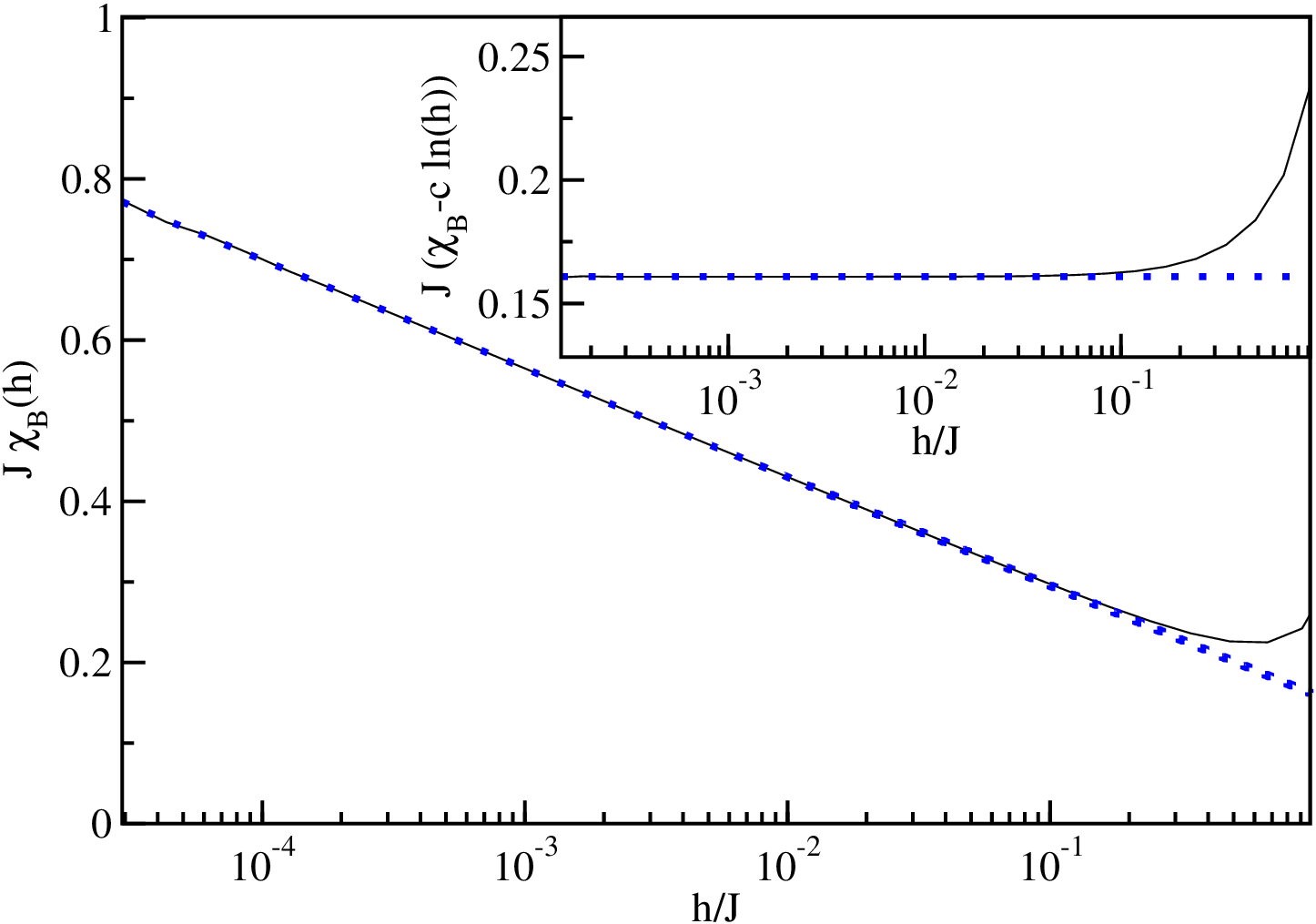}
\end{center}
\caption{Numerical data for the boundary susceptibility (black solid line) at $K=3/2$ (i.e.~$
  \g=\pi/3$) compared to formula (\ref{chi11}) (blue dotted line). The inset
  shows $\chi_B(h)-c \ln h$, where $c=-1/\sqrt 3 \pi^2$. According to
  Eq.~\refeq{chi11}, this yields the scale of the logarithmic divergence.}
\label{FigT0_2}
\end{figure}

\begin{figure}[!htp]
\begin{center}
\includegraphics*[width=0.75\columnwidth]{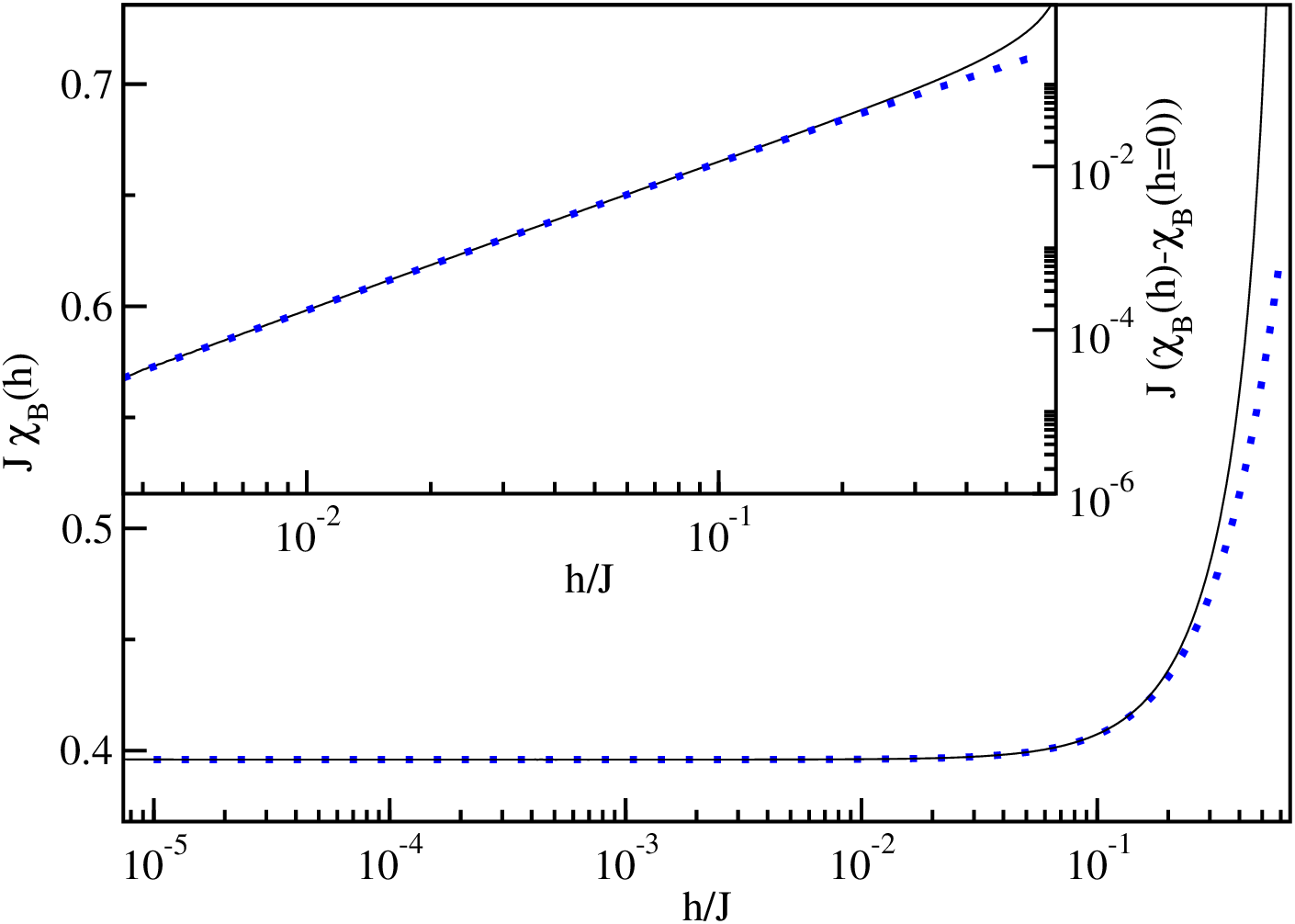}
\end{center}
\caption{Numerical data for the boundary susceptibility (black solid line) at $K=5/2$ (i.e. $\g=3\pi/5$) compared to
  formula (\ref{chi11}) (blue dotted line). For the plot in the inset, the constant $\chi_B(h=0)$ has been subtracted. Note the double logarithmic scale in the inset.}
\label{FigT0_3}
\end{figure}

\begin{figure}[!htp]
\begin{center}
\includegraphics*[width=0.99\columnwidth]{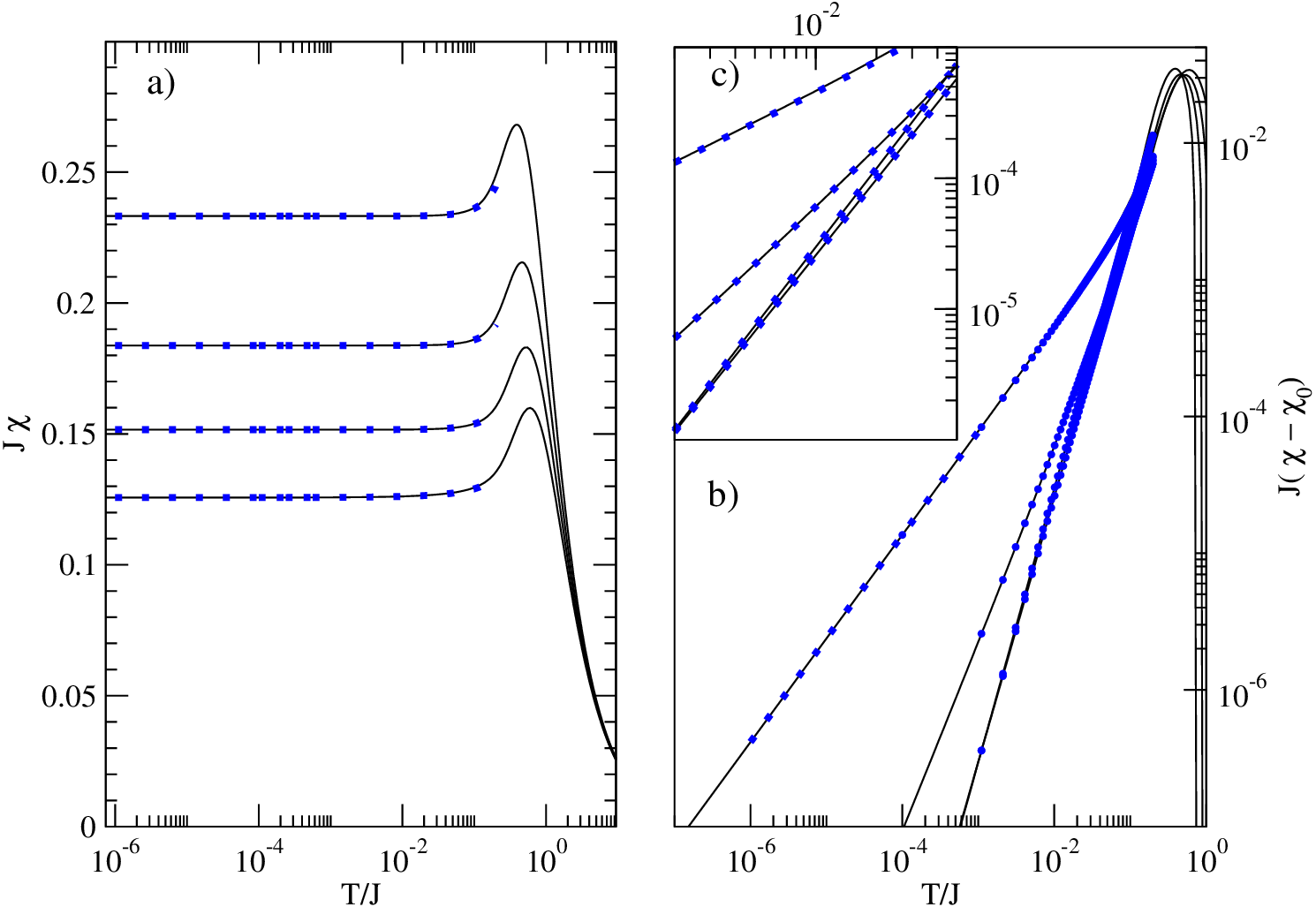}
\end{center}
\caption{Comparison between numerical data (black solid lines) for the bulk susceptibility in the
  repulsive regime and formula (\ref{B116}) (blue dotted lines) for
  $\g=0.5,0.8,\pi/3,1.3$, corresponding to $\Delta=0.88,0.7,1/2,0.27$ (from
  bottom to top in a) and from top to bottom in b), c)). In b), the constant
  $\chi(T=0,h=0)$ has been subtracted; the inset c) is a zoom of b). }
\label{Fig_fT1}
\end{figure}

\begin{figure}[!htp]
\begin{center}
\includegraphics*[width=0.99\columnwidth]{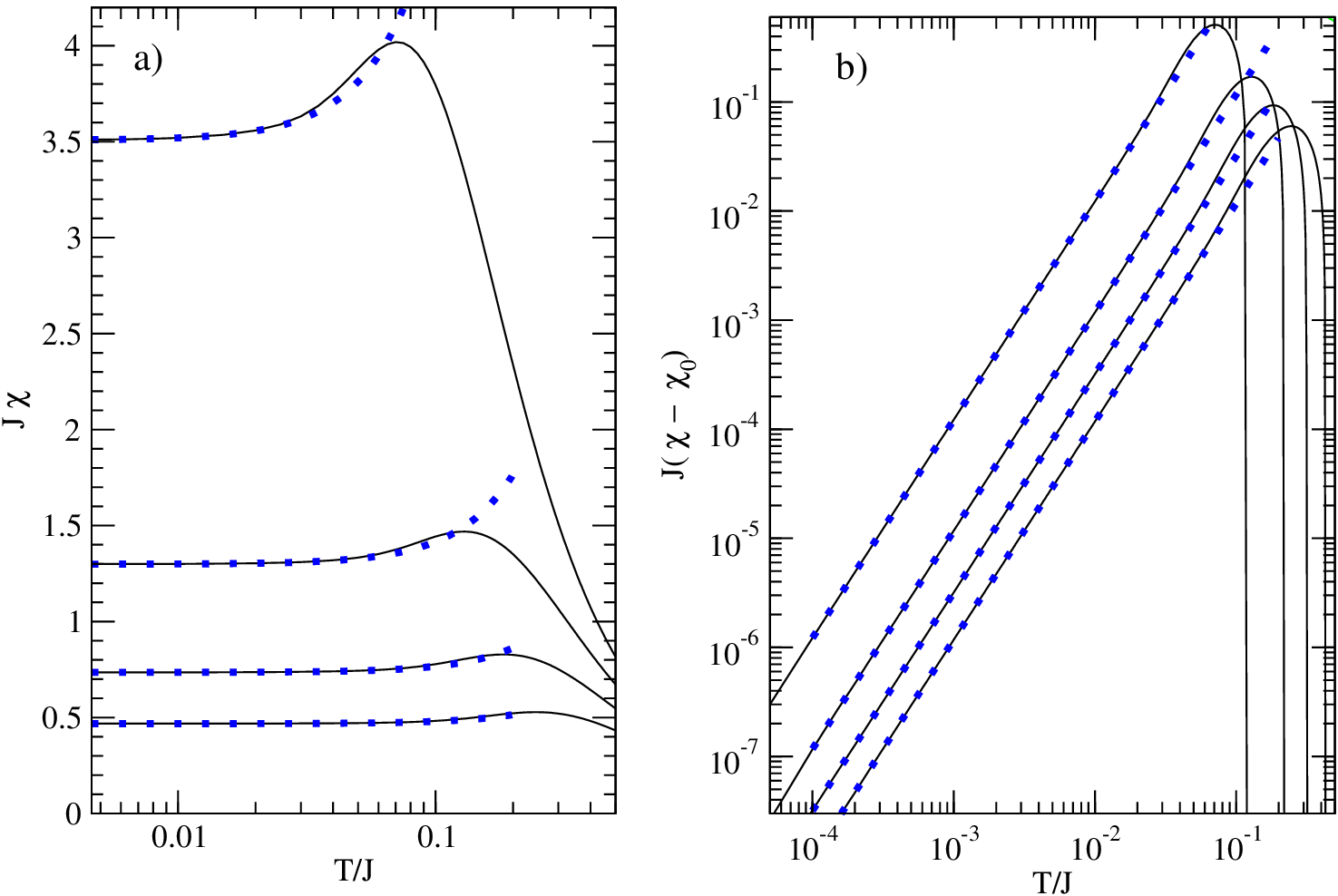}
\end{center}
\caption{Comparison between numerical data (black solid lines) for the bulk susceptibility in the
  attractive regime and formula (\ref{B116}) (blue dotted lines) for
  $\g=\pi-0.5, \pi-0.8, 2 \pi/3,\pi-1.3$, corresponding to
  $\Delta=-0.88,-0.7,-1/2,-0.27$ (from top to bottom in a) and b)). In b), the
  constant $\chi(T=0, h=0)$ has been subtracted.}
\label{Fig_fT2}
\end{figure}

To check our analytical formulas, in particular those for the boundary
susceptibility, we use numerical data. For $T=0$ we have solved
Eq.~(\ref{int1}) numerically (for more details see \cite{BortzSirker}). This
allows us to obtain $\chi_B(h)$ and to compare with the analytical formula
(\ref{chibbz}) if $|h|\ll \alpha$. Such a comparison has already been performed
in \cite{BortzSirker} in the repulsive regime $0\leq\Delta\leq 1$ verifying
our analytical result for the constant term and for the term which involves a
fractional power of $h$. These are the dominant contributions for $1<K<5/2$.
For $K>5/2$ the leading term is quadratic in $h$. To check also this term we
present in Fig.~\ref{FigT0_1} numerical data in the attractive regime.
Another point worth checking are the two anisotropies $K=3/2$ and $K=5/2$
where we have obtained the explicit formulas (\ref{chi11}) and (\ref{K5.2}) for
the logarithmic corrections. These are compared with numerical data in
Fig.~\ref{FigT0_2} and Fig.~\ref{FigT0_3}, respectively. 
Next, we turn to finite temperatures. The bulk quantities can be calculated
numerically based on the Bethe ansatz solution in the quantum-transfer-matrix
approach \cite{Kluemper_HB}. A comparison between these data and
Eq.~(\ref{B116}) in the repulsive and in the attractive regime is shown in
Fig.~\ref{Fig_fT1} and Fig.~\ref{Fig_fT2}, respectively. Figs.~\ref{Fig_fT1}b)
and \ref{Fig_fT2}b) show $\chi(T,h=0)-\chi(T=0,h=0)$ in double logarithmic
plots.  Excellent agreement is found for both the amplitudes and the exponents
of the leading $T$-corrections, thus confirming (\ref{B116}) first derived by
Lukyanov \cite{Lukyanov}. In particular, note the dependence of the exponent of
the leading $T$-correction on the anisotropy in the repulsive regime,
Fig.~\ref{Fig_fT1}b), whereas this exponent is constant in the attractive
regime, Fig.~\ref{Fig_fT2}b).

\begin{figure}[!htp]
\begin{center}
\includegraphics*[width=0.75\columnwidth]{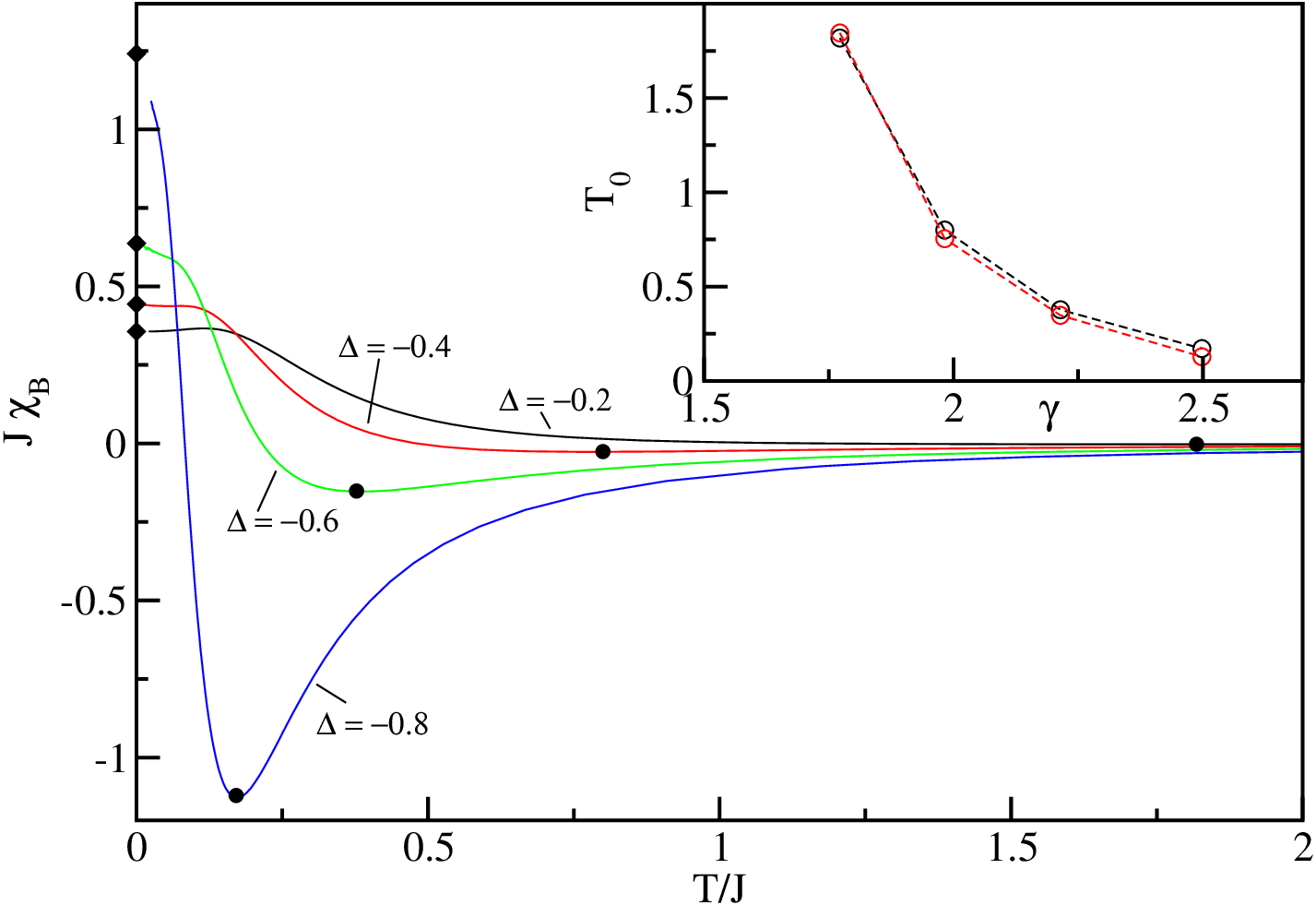}
\end{center}
\caption{TMRG data for $\chi_B(T)$ in the attractive regime. The black diamonds
  represent the zero temperature result known from BA. The black dots
  denote the minima $T_0$ of $\chi_B(T)$. In the inset we show that these minima
  (black dots) are well described by formula (\ref{N1}) (red dots) with
  $A=1.245$ obtained from a fit. The dashed lines are a guide to the eye.}
\label{Fig_TMRG1}
\end{figure}

For the boundary susceptibility at finite temperature no Bethe ansatz solution
is known today. We therefore have calculated this quantity by using the
density-matrix renormalisation group applied to transfer matrices (TMRG). This
method is particularly suited because the thermodynamic limit can be performed
exactly and no finite size corrections disguise the boundary contributions we
are looking for. This method has already been used in \cite{BortzSirker} to
obtain data in the repulsive regime and the reader is referred to this article
for more details about the TMRG. In Fig.~\ref{Fig_TMRG1} we show data for the
boundary susceptibility in the attractive regime.
Interestingly, $\chi_B$ first decreases when starting from infinite
temperature down to a temperature $T_0$ which depends on anisotropy before it
starts increasing at lower temperatures. This behaviour can be understood in
terms of a crossover from antiferromagnetic to ferromagnetic short-range order
in the bulk with increasing temperature, as studied in \cite{KluemperMcCoy}.
According to \cite{KluemperMcCoy}, a change in the dispersion relation of
elementary excitations from linear to quadratic behaviour occurs when
increasing the temperature from $T<T_0(\g)$ to $T>T_0(\g)$. Note that the
critical temperature $T_0(\g)$ depends on the anisotropy. 

We expect that the crossover temperature between these two regimes is
described by the same formula for $T_0(\g)$ as for the bulk, which reads
\cite{KluemperMcCoy}
 \begin{equation}
\label{N1}
T_0(\g) = A\frac{\sin\gamma}{\gamma}\tan\frac{\pi(\pi-\gamma)}{2\gamma}.
\end{equation}
Here $A$ is determined by a matrix element which is different for boundary and
bulk susceptibility. This matrix element cannot be calculated by BA. In the
inset of Fig.~\ref{Fig_TMRG1} we show that this formula indeed describes very
well the minima which occur in $\chi_B$ where $A$ is obtained from a fit.

Finally, we have a closer look at the low-temperature regime of
Fig.~\ref{Fig_TMRG1} and compare the TMRG data with (\ref{B119}).
\begin{figure}[!htp]
\begin{center}
\includegraphics*[width=0.75\columnwidth]{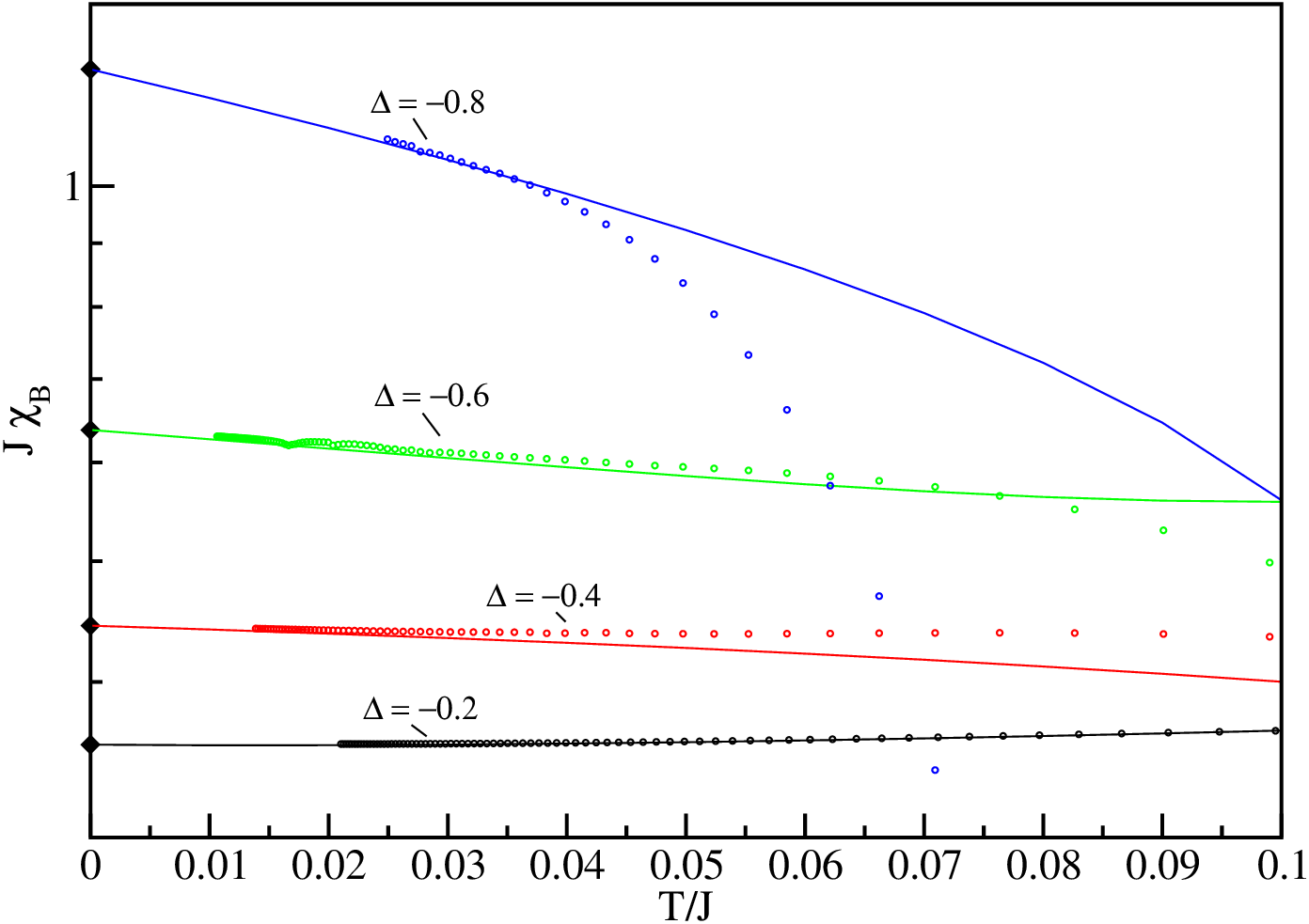}
\end{center}
\caption{Comparison between the TMRG data (circles) and (\ref{B119}) (solid
  lines) for $\chi_B(T)$ in the attractive regime. The black diamonds denote the $T=0$ BA result.}
\label{Fig_TMRG2}
\end{figure}
Fig.~\ref{Fig_TMRG2} shows excellent agreement confirming that $\chi_B$
depends linearly on temperature in the attractive regime for $T\ll 1$.  
\section{Conclusions}
\label{conclusions}
We have studied low-energy thermodynamic and ground-state properties of the
open spin-1/2 $XXZ$-chain by combining the bosonisation technique with the
exact Bethe ansatz solution in the critical regime $-1<\Delta<1$. Bosonisation
has allowed us to obtain a low-temperature, low-field expansion of the free
energy in terms of $T,h$ for both the bulk and the boundary parts of the free
energy with unknown coefficients. We have argued that the coefficients of the
regular terms in the boundary part (these are those powers of $h,T$ which
appear also in the bulk part) involve the short-distance lattice cutoff and
therefore cannot be obtained from field theory. All leading terms in the bulk
part, on the other hand, are cutoff independent and their coefficients could
be determined from Bethe ansatz by expanding the ground-state energy in terms
of the magnetic field $h$. For the boundary free energy only the coefficients
of two terms turned out to be cutoff independent.\footnote{There are also
  higher powers in $h,T$ which will be cutoff independent, but those terms are
  not very helpful because they are next-leading compared to a term with
  unknown coefficient.} These are a $h^2T$- and a $h^2T^{2K-3}$-term.
Fortunately, these two terms combined with a $h^2$-term obtained by BA give a
complete low-energy description of the boundary susceptibility $\chi_B$ for
all anisotropies.

We used the Wiener-Hopf procedure within the Bethe ansatz in the thermodynamic
limit to obtain the ground-state energy and a systematic low-field expansion
of the susceptibility at zero temperature. The coefficients of the leading and
next-leading terms in this expansion have been calculated explicitly for both
the bulk and the boundary contributions. The possible exponents of higher
order terms have been classified in a systematic way.

We compared our results with numerical data obtained by various techniques:
The $T=0$ Bethe ansatz equations involved solving linear integral equations
numerically, the finite-$T$-behaviour of the bulk susceptibility was
calculated in the quantum transfer matrix approach, and the
finite-$T$-behaviour of the boundary susceptibility was obtained from the
density-matrix renormalisation group applied to transfer matrices. In all
cases, we found excellent agreement between analytical and numerical results.

We identified several crossover phenomena. Depending on the anisotropy, the
crossover of scaling dimensions of the irrelevant operators in the low-energy
effective Hamiltonian leads to a crossover of critical exponents of $T, h$ in
the free energy. Associated with this crossover are divergences in the
corresponding amplitudes, which cancel to yield contributions logarithmic in
$T,h$ at the crossover points. Whereas this happens in the next-leading orders
in the bulk, this affects the boundary contributions both in leading and in
next-leading orders. Furthermore, we observed a crossover in the boundary
susceptibility in the attractive regime, due to competing antiferromagnetic or
ferromagnetic ordering tendencies at low and high temperatures, respectively.

We think that the rich results of the boundary behaviour analysed in this
article stimulate further research in direction of an exact treatment of the
boundary free energy at finite temperatures.

\section*{Acknowledgments}
We thank I. Affleck, V. Bazhanov, A. Kl\"umper and B. McCoy for stimulating
discussions. We also acknowledge support by the German Research Council ({\it
  Deutsche Forschungsgemeinschaft}) and are grateful for the computing
resources provided by the Westgrid Facility (Canada).

\end{document}